\documentclass[preprint,authoryear,12pt]{elsarticle}

\journal{Computational Statistics and Data Analysis}

\usepackage[update,prepend]{epstopdf}
\usepackage{mathrsfs}
\usepackage{verbatim}
\usepackage{amsfonts}
\usepackage{epsfig}
\usepackage{amsmath}
\usepackage{color,soul}
\usepackage[a4paper, portrait, margin=0.75in]{geometry}
\usepackage[usenames,dvipsnames]{xcolor}
\usepackage{float}
\usepackage{xcolor}
\usepackage[hyperindex,breaklinks]{hyperref}

\newcommand{\hlc}[2][yellow]{ {\sethlcolor{#1} \hl{#2}} }
\definecolor{light-gray}{gray}{0.8}

\begin{document}

\begin{frontmatter}

\title{Cross-Validated Wavelet Block Thresholding \protect \\ for Non-Gaussian Errors}
\author[ida]{K.˜McGinnity}
\author[fsu]{R.˜Varbanov\corref{cor1}}
\ead{r.varbanov@stat.fsu.edu}
\author[fsu]{E.˜Chicken}
\cortext[cor1]{Corresponding author}
\address[ida]{Institute for Defense Analyses, Alexandria, Virginia}
\address[fsu]{Department of Statistics, Florida State University, Tallahassee, FL}

\begin{abstract}
Wavelet thresholding generally assumes independent, identically distributed normal errors when estimating functions in a nonparametric regression setting. VisuShrink and SureShrink are just two of the many common thresholding methods based on this assumption. When the errors are not normally distributed, however, few methods have been proposed. A distribution-free method for thresholding wavelet coefficients in nonparametric regression is described, which unlike some other non-normal error thresholding methods, does not assume the form of the non-normal distribution is known. Improvements are made to an existing even-odd cross-validation method by employing block thresholding and level dependence. The efficiency of the proposed method on a variety of non-normal errors, including comparisons to existing wavelet threshold estimators, is shown on both simulated and real data.
\end{abstract}


\end{frontmatter}

\section{Introduction}
Wavelet thresholding has been a staple of statistical functional estimation for years. \cite{donoho4,  donoho3, donoho} introduced VisuShrink and SureShrink methods for thresholding the wavelet coefficients derived from the wavelet transformation of the observed data in nonparametric regression:
\begin{equation}
y_i=f(x_i) + \varepsilon_i, \hspace{.5cm} i= 1, 2, \ldots, n, \label{eq:regr}
\end{equation}
where the $\varepsilon_i$ are independent and identically distributed (iid) Gaussian errors with mean 0 and constant variance $\sigma^2$ and the sample points $x_i=i/n$ are equally spaced over an interval. The assumptions on the errors have been loosened in only a handful of papers on wavelet thresholding.

\cite{neumann} discuss wavelet thresholding methods in non-Gaussian and non-iid situations. The main idea of their paper is that, in many situations, asymptotic normality can be proven and traditional thresholding methods can be used. Given independent observations, they demonstrate a way to show equivalence to the Gaussian case via strong approximations. They also derive asymptotic normality in the case of weak dependence. 

\cite{anton} propose a scale-dependent wavelet thresholding procedure for Gaussian noise, and then extend it to include non-Gaussian noise. However, the paper assumes not only that the non-normal errors are iid with mean zero, but also that they follow a known specified distribution. They determine a suitable threshold for each resolution level by mimicking the arguments of Donoho and Johnstone in the Gaussian case.

\cite{pensky} investigate the performance of Bayes factor estimators in wavelet regression models with iid non-Gaussian errors. They choose a general distribution $\eta_j$ for the errors and assume they possess symmetric PDFs on the reals that are unimodal, positive, and finite at zero. One advantage of their method is that knowledge of the true distribution of the errors is not needed in order to obtain an optimal estimator of $f$. However, their estimators are only preferable for irregular functions with high peaks, and produce sub-optimal results when compared with other methods under certain prior distributions.

\cite{nason2} introduces an even-odd cross-validation method for choosing the threshold parameter in wavelet shrinkage. His statistic compares an interpolated wavelet estimator from the even reconstructed data to the odd noisy data and vice versa over various threshold values, then applies a sample size correction.

In this paper, we propose a completely nonparametric method to threshold wavelet coefficients that enhances Nason's cross-validation method by incorporating level-dependent block thresholding. Block thresholding thresholds wavelet coefficients in groups, rather than individually, with the goal of increasing precision by utilizing information about neighboring coefficients (\cite{cai}). Nason's method uses term-by-term thresholding, so it is reasonable to ask if incorporating blocking will have an analogous effect here. 

Nason also makes use of a global threshold, the same threshold value for all considered coefficients. This is similar to VisuShrink of \cite{donoho4}. However, level-dependent thresholding has also been shown to have advantages over universal thresholds. For example, SureShrink (\cite{donoho3}), a level-dependent thresholding method, has been shown to have lower MSE than VisuShrink. Each of these modifications, blocking and level dependence, improves performance with distribution-based thresholds and thus are natural considerations for attempting to improve cross-validation thresholding.

Our method does not put any assumptions on the errors except that they are iid and centered at zero. Unlike Neumann and von Sachs, we do not discuss asymptotic normality, but instead develop a method specifically meant to handle non-Gaussian errors. Nor do we require that the distribution of the errors be known, as do Antoniadis and Fryzlewicz. Unlike that of Pensky and Sapatinas, no proper choice of prior is required for our method. 

This paper is divided as follows. Section 2 provides a brief background on wavelets, wavelet notation, and wavelet thresholding methods before the details of the proposed estimator are described in Section 3. Section 4 contains a simulation comparison of the proposed estimator to the Nason estimator, VisuShrink, and other current methods which assume normal errors, as well as two example applications of our method using inductance plethysmography and vertical profile density data. A discussion of the results and methods is given in the final section.

\section{Background}

\subsection{Wavelets}

Wavelets are an orthogonal series representation of functions in the space of square-integrable functions $L_2(\mathbb{R})$. \cite{ogden} and \cite{vidakovic} offer good introductions to wavelet methods and their properties. Let $\phi$ and $\psi$ represent the father and mother wavelet functions, respectively. There are many choices for these two functions, see \cite{daubechies}. Here, $\phi$ and $\psi$ are chosen to be compactly supported and to generate an orthonormal basis. Let
$$
\phi_{jk}(x) = 2^{j/2}\phi(2^{j}x-k)
$$
and
$$
\psi_{jk}(x)  = 2^{j/2}\psi(2^{j}x-k)
$$
be the translations and dilations of $\phi$ and $\psi$, respectively. For any fixed integer $j_0$,
$$
\{\phi_{j_0k},\psi_{jk}|j\geq j_0, k \mbox{ an integer}\}
$$
is an orthonormal basis for $L_2(\mathbb{R})$. Let
$$
\xi_{jk}=\langle f,\phi_{jk} \rangle
$$
and
$$\theta_{jk}=\langle f,\psi_{jk} \rangle
$$
be the usual inner product of a function $f\in L_2(\mathbb{R})$ with the wavelet basis functions. Then $f$ can be expressed as an infinite series:
\begin{equation}
f(x)=\sum_k\xi_{j_0k}\phi_{j_0k}(x)+\sum_{j=j_0}^{\infty}\sum_k\theta_{jk}\psi_{jk}(x).
\label{eq:series}
\end{equation}
The function $f$ is not known and must be estimated. This is done using the discrete wavelet transform (DWT) of \cite{mallat}. If $f$ is sampled as a vector of dyadic length $n=2^J$ for some positive integer $J$, then the DWT will provide a total of $n$ estimated coefficients $\xi_{j_0k}$ and $\theta_{jk}$ over the indices $j=j_0,j_0+1,\ldots,J-1$ and for all appropriate $k$. The lowest level possible for $j_0$ is 0, the highest is $J-1$.

The wavelet basis functions are easily periodized to a specified interval. In this paper, we use wavelets that have been periodized to the interval $[0, 1]$. In this case, the index $k$ for resolution level $j$ runs from $1$ to $2^j$ in (\ref{eq:series}).

Wavelets have the useful property that they can simultaneously analyze a function in both time and frequency. This is done by projecting the function to be analyzed into several subspaces or resolution levels. Each resolution level represents a different degree of smoothness of the function. The lowest resolution level, associated with the index $j=j_0$, represents the smoothest or coarsest part of the function. Increasing the index $j$ corresponds to decreasing smoothness. The highest resolution levels $j$ therefore represent the behavior of the function at the highest frequencies or scales. Since the wavelet series (\ref{eq:series}) forms an orthogonal representation, the sum of the projections in these resolution levels is the original function $f$. The construction of wavelet functions $\phi$ and $\psi$ provide the ability to localize the analysis within each subspace. The higher the resolution, the greater the degree of localization.

By varying the resolution level $j$, wavelets have the ability to zoom in or out onto the smooth or detailed structure of $f$. This is referred to as the multiresolution property of wavelets. Changing the index $k$ allows wavelets to localize the analysis. These properties enable wavelets to model functions of very irregular types, as well as smooth functions.

We use $W$ to denote the $n\times n$ DWT transformation matrix. Applying the DWT to the observed values $y=(y_1, y_2, \ldots, y_n)^\prime$ in (\ref{eq:regr}) gives the estimated wavelet coefficients
$$
\tilde\theta=
\left(\tilde\xi_{j_01}, \tilde\xi_{j_02}, \ldots, \tilde\xi_{j_02^{j_0}}, \tilde\theta_{j_01}, \tilde\theta_{j_02}, \ldots, \tilde\theta_{J-1,2^{J-1}}\right)^\prime
=Wy.
$$
Applying the inverse DWT $W^{-1}=W^\prime$ to these coefficients returns the original data, $y=W^\prime Wy$.

\subsection{Thresholding}
Most wavelet analysis uses some form of thresholding. These can be term-by-term methods, where each individual wavelet coefficient is modified individually, or block methods, where coefficients are modified in groups. There are two general types of term-by-term thresholding methods. Hard thresholding is a ``keep or kill'' operation that sets to $0$ any wavelet coefficient below a certain threshold value $\lambda$. In soft thresholding, coefficients smaller in magnitude than $\lambda$ are set to $0$ and coefficients larger than $\lambda$ are shrunk towards $0$.

A popular method of term-by-term thresholding is the VisuShrink method of \cite{donoho4}. For a single estimated coefficient $\tilde\theta_{jk}$,
$$
\hat\theta_{jk}=\eta(\tilde\theta_{jk}, \lambda)= \mbox{sgn}(\tilde{\theta}_{jk})(|\tilde{\theta}_{jk}|-\lambda)_+
$$
is the thresholded coefficient, where $\lambda$ is a threshold parameter. This is an example of soft thresholding. Hard thresholding via Visushrink can also be used. The value of $\lambda$ is chosen to give optimal results in terms of reconstruction; $\lambda=\sigma\sqrt{2\log(n)}$ where $\sigma$ and $n$ are from (\ref{eq:regr}). This is known as the universal threshold since the same threshold value is used across all resolution levels. A thresholded estimate is then formed as
$$
\hat{f}=W^\prime\eta(Wy, \lambda).
$$
Note that only select resolution levels are subjected to thresholding. Typically these are the highest $m$ resolution levels, where $m$ is specified by the user.

Another popular thresholding method combines VisuShrink with a resolution-level-dependent method based on Stein's Unbiased Risk Estimator (SURE), see \cite{stein}. This SURE based method obtains optimal mean squared error (MSE) but does not work well if the detail coefficients at a given resolution level are too sparse. Thus, this hybrid method will implement the universal threshold if a level is too sparse, and will use the SURE based method otherwise. Level $j$ is too sparse if 
$$
2^{-j} \displaystyle\sum\limits_{k=1}^{2^j} (d^2_{j,k}/\sigma^2_{j,k} - 1) \leq 2^{-j/2} (\log_2 2^j)^{3/2}.
$$

One may choose to ignore the sparsity condition. We will refer to this as SureShrink in our paper, see \cite{donoho3}. Both the hybrid method (HybridShrink) and SureShrink are only designed to be used with soft thresholding, and are more computationally expensive than VisuShrink.

Block thresholding methods typically employ variants of the block projection estimator of \cite{cai}. If $B_i$, $i=1,...,N$ are blocks of equal size $L$ ($\sim \log n$) that evenly divide the observed wavelet coefficients $\tilde{\theta}$, $\mathscr{H}$ is a subset of the block indices $\{1,...,N\}$, and $\tilde{\theta}_{B_i}$ are the $L$ coefficients in block $B_i$, the goal is to have $\mathscr{H}$ consist only of those blocks where the signal is greater than the noise: 

\begin{equation}
\mathscr{H}  = \mathscr{H}(\tilde{\theta}) = \lbrace i: \|\tilde{\theta}_{B_i}\|^2_2 > L\sigma^2 \rbrace,
\end{equation}
where $\|\tilde{\theta}_{B_i}\|^2_2 = \sum_{k \in B_i} \tilde{\theta}_{k}^2.$

Anything less than the threshold is set to zero, giving the final block projection estimator to be

\begin{equation}
\hat{\theta}_{B_i} = 
\left\{
\begin{array}{cc}
	\tilde{\theta}_{B_i}	&	\text{if } i \in \mathscr{H} \\ 
	0	&	\text{if } i \not \in \mathscr{H}.	\\
\end{array}
\right.
\end{equation}

So, in any particular block, either the original coefficients are kept, or every coefficient in the block is set to zero.

In more recent advances, \cite{barber} build on the multiwavelet shrinkage method introduced by \cite{Downie96thediscrete} by implementing the use of a complex wavelet transform to estimate real signals. The Complex Multiwavelet Style (CMWS) shrinkage method uses the complex-valued Daubechies wavelets used by \cite{Lawton} and \cite{Lina} and discards the imaginary component of the reconstructed signal. For complex-valued empirical wavelet coefficient $ d_{j,k}^* $, where $ d_{j,k}^* \sim N_2(d_{j,k},\Sigma_j) $, a ``thresholding statistic'' \begin{equation} \theta_{j,k} = d_{j,k}^{*T}\Sigma_j^{-1}d_{j,k}^* \end{equation} is computed. Hard thresholding is then carried out according to the rule \begin{equation} \widehat{d}_{j,k}^{MH} = d_{j,k}^*I(\theta_{j,k}>\lambda), \end{equation} where \textit{I} is the indicator function and $ \lambda = 2 \log n $, a theshold derived by \cite{Downie96thediscrete} for use in the multiwavelet case.

\cite{fryzlewicz} combines thresholding and the unbalanced Haar (UH) basis introduced by \cite{Girardi}, for which jumps in the basis functions do not necessarily occur in the middle of their support. Each basis vector for the discrete UH transform is selected to best match the data at a specific scale and location by applying the idea of a \textit{matching pursuit} algorithm originally introduced by \cite{mallat3}. The author focuses on a \textit{top-down} version of algorithm where the greatest concentration of the power of signal is placed on the coarse scales before proceeding to finer scales. After transformation, hard thresholding can be carried out by comparing the absolute value of empirical wavelet coefficient $ Y_{j,k} $ to the universal threshold used in Visushrink, $ \lambda = \sigma \sqrt{2 \log n} $.

\cite{johnstone2} propose the selection of a level-dependent threshold utilizing Empirical Bayes methods. Acknowledging the need for a threshold to adapt to the sparsity of a signal and consequently the resolution level, this Empirical Bayes method models this sparsity by selecting a suitable prior distribution for the wavelet coefficients of the signal. At each resolution level, the weights of the distributions that comprise the mixture prior are estimated using marginal-maximum likelihood. A threshold is then calculated such that the posterior median of the distribution of the coefficients is zero if and only if the magnitude of the coefficients is less than said threshold. Hard thresholding can then be applied using the selected threshold level. 

It is important to note that all the methods mentioned above put strong assumptions on the errors, particularly normality.

The properties and advantages of differing types of thresholding methods are well documented in the literature. See \cite{donoho4, donoho, cai, pensky2, caisilverman, chicken3, chicken4}.

\section{The Estimator}

We observe a vector of data, $y_i = f(x_i) + \epsilon_i$,  $i = 1,...,n$ for which the noise $\epsilon_i$ is independent and identically distributed. If the noise is not centered at zero, we subtract the mean of the noise from each $\epsilon_i$ to eliminate any potential bias. Specific choices for the noise will be discussed in more detail later. When the discrete wavelet transform (DWT) is applied to the data, we obtain $Wy = Wf(x) + W\epsilon$ where $W$ represents the wavelet transformation. If the original errors $\epsilon$ have mean $0$ and covariance $\Sigma=\sigma^2I_n$ then,

\begin{equation*}
E(W\epsilon) = 0	
\end{equation*}
and
\begin{equation*}
\text{Cov}(W\epsilon) = W\Sigma W^\prime = W\sigma^2 I_n W^\prime =\sigma^2 W I_n W^\prime =\sigma^2 W W^\prime	=\sigma^2 I_n.
\label{eq:cov}
\end{equation*}

Since $W$ is an orthogonal transformation, these new errors $W\epsilon = \epsilon^{\prime}$ are uncorrelated, as shown above by the fact that the covariance matrix is diagonal. They are not necessarily independent unless the $\epsilon_i$ have a Gaussian distribution. \cite{opsomer} show that issues in nonparametric regression only arise with \textit{correlated} errors. The errors are also identically distributed within each resolution level (see \cite{M1006}).

Our approach is motivated by Nason's even-odd cross-validation procedure. We propose three major modifications. First, we make the threshold level-dependent, meaning we allow a different threshold value to be chosen for each resolution level, rather than a global threshold for all resolution levels. This provides increased estimability and accuracy. For example, SureShrink, which has a level-dependent threshold, has lower MSE than VisuShrink, which has a global threshold, but SureShrink often has noisier reconstructions. Second, we employ block thresholding. Block thresholding divides the data into blocks of neighboring coefficients, and will set to zero an entire block if the sum of the squared coefficients in that block is less than the threshold value. Block thresholding has been shown to have improved error and visual fit over traditional term-by-term thresholding (\cite{cai}), alleviating the above concern about ``visually pleasing'' reconstructions vs.\ low MSE. Third, we add an additional term to the error statistic that gets minimized during cross-validation. This term compares two sampled reconstructions to one another.

To set up our algorithm, we first divide the noisy data of length $n$ into even and odd parts, $y^e$ and $y^o$:

\begin{equation*}
y^e = \left(y_2, y_4,..., y_n\right)^\prime	
\end{equation*}
and
\begin{equation*}
y^o = \left(y_1, y_3,..., y_{n-1}\right)^\prime	.
\end{equation*}

\noindent We then apply the DWT to each part to obtain $\hat{f}^{e}$ and $\hat{f}^{o}$. We choose the block size to be the closest dyadic number to $\log(n)$ (shown to possess optimal properties in \cite{cai}). We define the threshold search range to be zero to the maximum sum of squared coefficients in any block. This is the most conservative search range we can choose since the minimum threshold will set all detail coefficients to be zero, giving the smoothest possible reconstruction, and the maximum threshold will set nothing to zero and return the original noisy data. It is possible to reduce the size of this search range through methods such as pilot density estimation on coefficients in the first resolution level, however, in doing so, one risks leaving out the optimal threshold value from the search. 

In the first iteration, we start by finding the optimal threshold value $\lambda$ for all detail resolution levels simultaneously by minimizing a combination of two errors. The first error is the sum of the squared difference between the odd reconstruction and the even data, and the squared difference between the even reconstruction and the odd data. This mimics the method of \cite{nason}, except we apply the threshold rule to all coefficients in a block rather than individual coefficients. The second error is the sum of the squared difference between the odd reconstruction and the even reconstruction. This error is not considered by Nason. We choose to use this combination of errors in order to obtain the optimal smoothness; intuitively, we expect a comparison of the noisy data to the smooth reconstruction to under-smooth our final estimate and a comparison solely between the even and odd reconstructions to over-smooth our final estimate. Thus, the initial global threshold estimate is  
\begin{equation}
 \hat{\lambda} = \underset{\lambda}{\operatorname{argmin}} \left\lbrace\frac{1}{2}\{\|\hat{f}^{o}_{\lambda} - y^{e}\|^2_2 + \|			\hat{f}^{e}_{\lambda} - y^{o}\|^2_2\} +  \frac{1}{2}\{\|\hat{f}^{o}_{\lambda} - \hat{f}^{e}_{\lambda}\|^2_2\}\right\rbrace. \label{eq:lambdahat}
\end{equation}

We then proceed to make our estimator level-dependent. Letting $\lambda_j$ be the threshold value for resolution level $j$ ($j=j_0,\ldots , J-1$), we fix $\lambda_{J-1}$ at the value found in (\ref{eq:lambdahat}) and repeat the search described above to find an optimal threshold for resolution levels $J-2$,\ldots ,$j_0$. This value becomes our $\lambda_{J-2}$ and we next search for the optimal threshold for resolutions levels $J-3$,\ldots ,$j_0$. This value becomes our $\lambda_{J-3}$ and we then find the optimal threshold for resolution level $J-4$,\ldots ,$j_0$. Repeat this process until the optimal $\lambda_{j_0}$ for $j_0$ alone is found. These serve as our initial level-dependent threshold estimates.

In the second and all subsequent iterations, we modify each resolution level threshold value one at a time until convergence is achieved. First, fix $\lambda_{J-2}$, $\lambda_{J-3}$,\ldots ,$\lambda_{j_0}$ at their previous iteration values and search for the optimal $\lambda_{J-1}$. Next, fix $\lambda_{J-1}$ at this new value and $\lambda_{J-3}$,\ldots ,$\lambda_{j_0}$ at their previous iteration values and search for the optimal $\lambda_{J-2}$. Then, fix $\lambda_{J-1}$ and $\lambda_{J-2}$ at their new values and $\lambda_{J-4}$,\ldots ,$\lambda_{j_0}$ at their previous iteration values and search for the optimal $\lambda_{J-3}$. Continue until all $\lambda_j$ have been modified individually. We found that this process generally only had to be performed once before all threshold values converged.

The last step in our procedure is also a modification of Nason's final step. Since all estimates in his paper were based on $n/2$ data points, a heuristic method based on the universal threshold was used to obtain a threshold for $n$ data points. Once his error statistic had been minimized to obtain a threshold value $\lambda^{n/2}$, the following correction was applied to obtain the final threshold:

\begin{equation}
\lambda^n \approx \left(1 - \dfrac{\log 2}{\log n}\right)^{-1/2}\cdot \lambda^{n/2}. \label{eq:nasoncorr}
\end{equation}
where the superscript refers to the number of points used to obtain the threshold value.

In our case, the sample size used to calculate the statistic changes depending on which resolution level $j$ we are working with ($j=1$ representing the highest resolution level). Thus we apply the following level-dependent correction:

\begin{equation}
\lambda_j^{n/2^j} \approx \left(1 - \dfrac{\log2}{\log(\frac{n}{2^{j}})}\right)^{-1}\cdot\lambda_j^{n/2^{j+1}},\  j = 1, 2, 3, 4,...,J-j_0 \label{eq:mycorr}
\end{equation}

Notice that we no longer need to take the square root in our correction since our threshold values are based on the sum of \textit{squared} coefficients within a block. Again, the superscript of $\lambda$ refers to the number points used to obtain the threshold value and does not indicate an exponent.

\section{Simulation and Examples}

\subsection{Simulation}

Our estimator is compared to the traditional thresholding method of VisuShrink, Nason's method, the CMWS shrinkage of \cite{barber}, the UH technique of \cite{fryzlewicz}, and the Empirical Bayes (EBayes) selection of thresholds of \cite{johnstone2} via mean squared error (MSE). (See \cite{M1006} report for comparisons to Antoniadis' method.)

\begin{figure}[H]
\begin{center}
\includegraphics[width=4in, height=3in, angle=0]{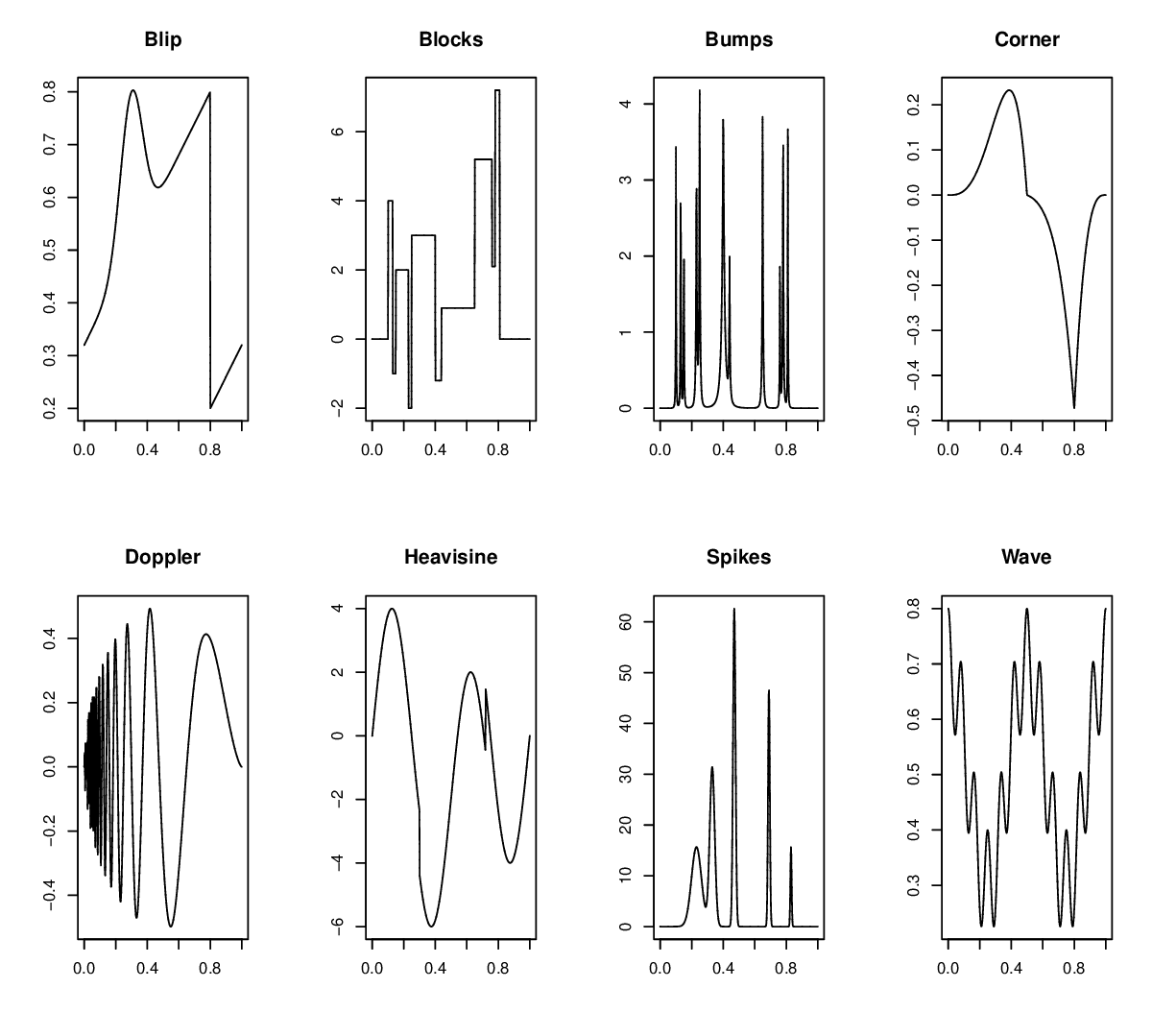}\\
\caption{The eight test functions used in this paper sampled at 512 points.}
\label{figure:TestFuncsClean}
\end{center}
\end{figure}

Using sample sizes ranging from $2^9$ to $2^{11}$ and signal to noise ratios (SNRs) of 3 and 5, these methods were tested on the eight standard test functions (Blip, Blocks, Bumps, Corner, Doppler, Heavisine, Spikes, and Wave), see \cite{donoho4} and \cite{marron}. Figure \ref{figure:TestFuncsClean} shows a plot of these functions. 

\begin{figure}[H]
\begin{center}
\includegraphics[width=4in, height=3in, angle=0]{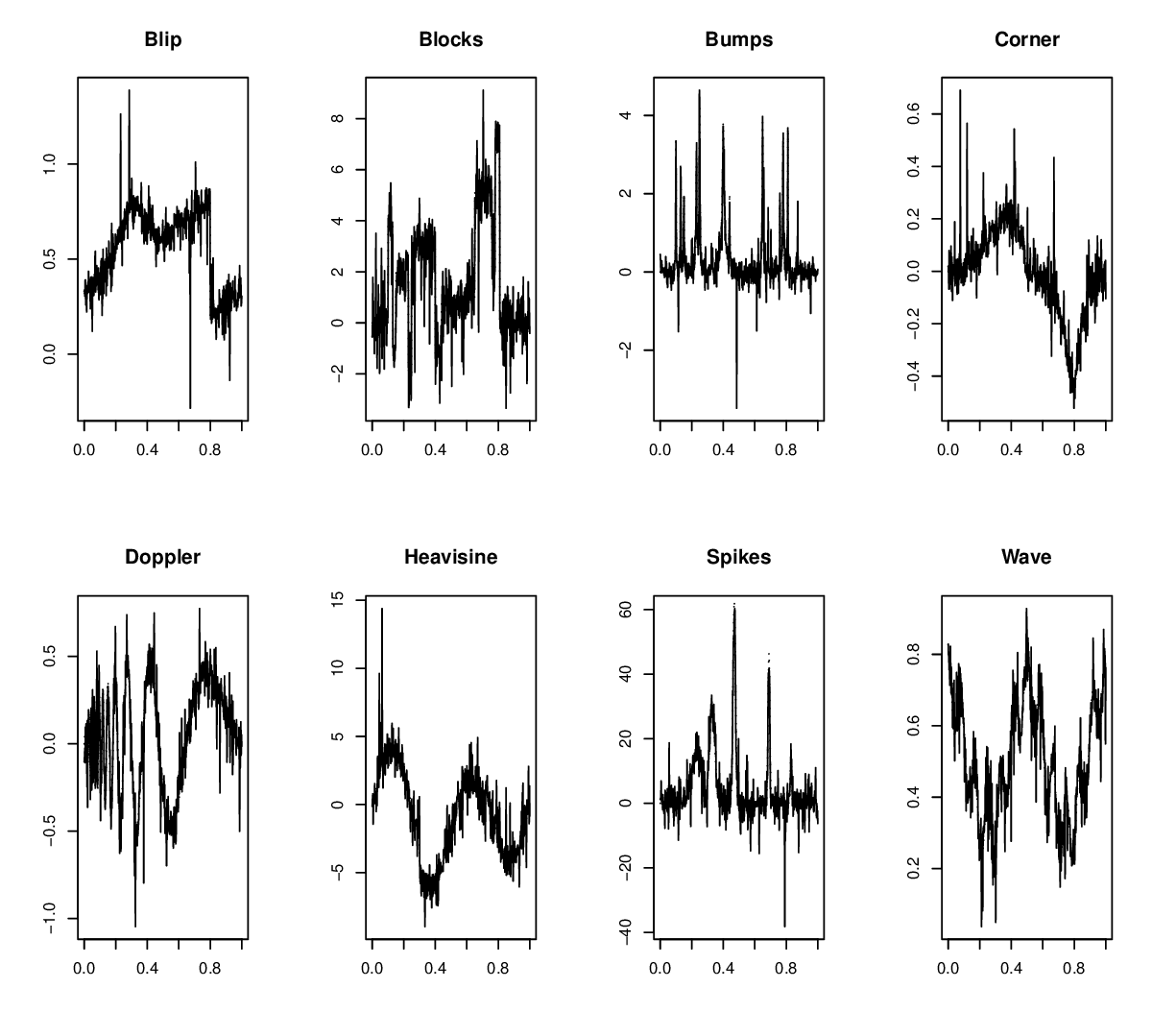}\\
\caption{The test functions with scaled $T_3$ noise added to represent a SNR of 5.}
\label{figure:TestFuncsT3}
\end{center}
\end{figure}

\begin{figure}[H]
\begin{center}
\includegraphics[width=4in, height=3in, angle=0]{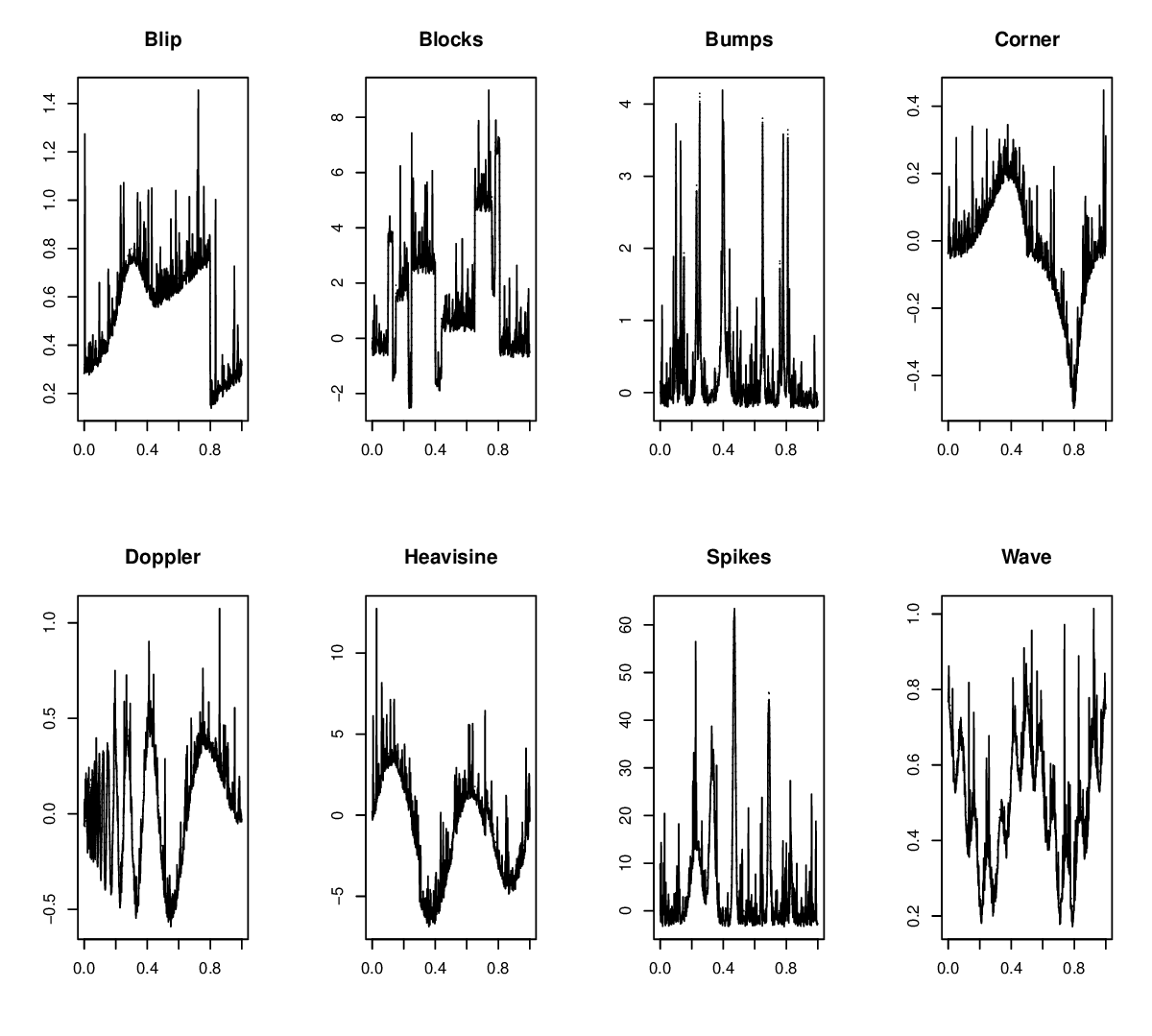}\\
\caption{The test functions with centered lognormal noise added to represent a SNR of 5.}
\label{figure:TestFuncsLognormal}
\end{center}
\end{figure}

Various types of noise were added on top of these test functions to create our data. The only assumptions imposed on the errors were that they were independent and identically distributed. In this paper we show results for $T_3$ and lognormal errors. Other types of noise were considered, but these two were chosen to exemplify a heavy-tailed distribution and a skewed distribution. The noise was scaled appropriately to obtain the desired SNR, thus the true distribution of the noise was not necessarily known. For example, a scaled $t$-distribution is no longer a $t$. In the case of skewed distributions such as lognormal, we first centered the noise at the mean before applying it to the test function. Figures \ref{figure:TestFuncsT3} and \ref{figure:TestFuncsLognormal} show plots of $T_3$ and lognormal noise added to all the test functions.

When possible, all functions were analyzed using the least asymmetric wavelet basis of length eight (LA-8) (\cite{daubechies}) with periodic boundary handling, and the lowest resolution level, $j_0$, set to $J-4$. For the CMWS method, estimation was done with all available complex-valued wavelets and the results were averaged. For the UH method, the functions were analyzed with the best top-down UH decomposition. For the EBayes method, a Laplace prior was used for the wavelet coefficients at each level. All simulations were performed in R (\cite{R}) using the \textit{waveslim} (\cite{R:waveslim}), \textit{WaveThresh} (\cite{R:wavethresh}), \textit{unbalhaar} (\cite{R:unbalhaar}), and \textit{EBayesThresh} (\cite{R:EBayesThresh}) packages. 

Tables \ref{table:T3_5}, \ref{table:T3_3}, \ref{table:Lognormal_5} and \ref{table:Lognormal_3} show simulation results of 100 repetitions each for $T_3$ and lognormal noise respectively. Hard thresholding and various combinations of $n$, SNR, and test function were used. The numbers in the table represent the ratio of the average MSE for the given method to the average MSE of VisuShrink. Thus a ratio less than one indicates that a particular method outperformed VisuShrink, while a ratio greater than 1 indicates that VisuShrink performed better. In parentheses, the standard deviation ($\times10^{-3}$) of the MSE from the 100 repetitions of the corresponding method is given. For each combination of conditions, we have identified the best performing method by highlighting the lowest average MSE. We also highlighted any other average MSE that was not significantly different than the lowest average MSE according to a paired \textit{t}-test at the $\alpha = 0.05$ significance level.

\begin{table}[H]
\caption{$T_3$ noise with SNR=5: Ratio of average MSE of listed method to average MSE of VisuShrink with corresponding standard deviation ($\times10^{-3}$) of MSE of listed method}
\label{table:T3_5}
\begin{center}
\begin{tabular}{ccccccc}
\hline
Function & n & LD Block & Nason & CMWS & UH & EBayes \\ 
\hline
 & 512 & 0.60(0.02)  & \hlc[light-gray]{0.47(0.05)}  & 0.94(0.21)  & 1.22(0.22)  & 1.35(0.23) \\
Blip  & 1024 & \hlc[light-gray]{0.43(0.02)}  & 0.49(0.04)  & 0.90(0.14)  & 1.14(0.15)  & 1.40(0.16) \\
 & 2048 & \hlc[light-gray]{0.29(0.02)}  & 0.36(0.08)  & 0.91(0.14)  & 1.10(0.15)  & 1.44(0.15) \\
\hline
 & 512 & 1.45(19.32)  & 1.10(15.92)  & 0.79(24.05)  & \hlc[light-gray]{0.61(25.04)}  & 1.08(25.95) \\
Blocks  & 1024 & 0.88(2.99)  & 0.85(5.92)  & 0.82(16.44)  & \hlc[light-gray]{0.71(17.11)}  & 1.21(18.52) \\
 & 2048 & \hlc[light-gray]{0.57(2.42)}  & 0.64(9.15)  & 0.82(16.19)  & 0.78(16.71)  & 1.28(17.36) \\
\hline
 & 512 & 1.68(2.04)  & 2.09(3.70)  & \hlc[light-gray]{0.65(2.37)}  & 0.92(2.45)  & 0.99(2.54) \\
Bumps  & 1024 & 1.33(1.45)  & 2.12(3.47)  & \hlc[light-gray]{0.76(1.65)}  & 1.18(1.76)  & 1.23(1.82) \\
 & 2048 & \hlc[light-gray]{0.84(0.72)}  & 1.18(1.90)  & \hlc[light-gray]{0.83(1.64)}  & 1.27(1.67)  & 1.33(1.73) \\
\hline
 & 512 & \hlc[light-gray]{0.20(0.01)}  & 0.25(0.03)  & 0.93(0.14)  & 1.35(0.14)  & 1.34(0.16) \\
Corner  & 1024 & \hlc[light-gray]{0.19(0.01)}  & 0.24(0.03)  & 0.92(0.09)  & 1.29(0.10)  & 1.41(0.11) \\
 & 2048 & \hlc[light-gray]{0.19(0.01)}  & 0.25(0.05)  & 0.92(0.10)  & 1.18(0.10)  & 1.44(0.10) \\
\hline
 & 512 & 0.91(0.24)  & 0.83(0.22)  & \hlc[light-gray]{0.74(0.46)}  & 1.50(0.49)  & 1.19(0.49) \\
Doppler  & 1024 & \hlc[light-gray]{0.42(0.11)}  & 0.73(0.14)  & 0.83(0.31)  & 1.53(0.32)  & 1.34(0.35) \\
 & 2048 & \hlc[light-gray]{0.43(0.03)}  & \hlc[light-gray]{0.47(0.17)}  & 0.87(0.31)  & 1.41(0.31)  & 1.41(0.33) \\
\hline
 & 512 & \hlc[light-gray]{0.21(4.28)}  & 0.26(9.32)  & 0.94(48.41)  & 1.44(50.53)  & 1.35(53.36) \\
Heavisine  & 1024 & \hlc[light-gray]{0.22(2.82)}  & 0.26(8.60)  & 0.92(32.64)  & 1.33(34.44)  & 1.40(38.01) \\
 & 2048 & \hlc[light-gray]{0.20(2.77)}  & 0.26(18.66)  & 0.92(32.66)  & 1.21(33.48)  & 1.44(35.04) \\
\hline
 & 512 & 1.09(345.62)  & 0.98(514.51)  & \hlc[light-gray]{0.81(555.43)}  & 1.20(571.63)  & 1.24(601.57) \\
Spikes  & 1024 & \hlc[light-gray]{0.46(111.53)}  & 0.56(145.44)  & 0.88(371.55)  & 1.31(386.95)  & 1.39(427.43) \\
 & 2048 & \hlc[light-gray]{0.22(37.16)}  & \hlc[light-gray]{0.27(206.09)}  & 0.90(370.63)  & 1.25(373.10)  & 1.43(398.18) \\
\hline
 & 512 & \hlc[light-gray]{0.46(0.03)}  & \hlc[light-gray]{0.47(0.02)}  & 0.75(0.14)  & 1.39(0.15)  & 1.07(0.15) \\
Wave  & 1024 & \hlc[light-gray]{0.21(0.01)}  & 0.25(0.03)  & 0.92(0.09)  & 1.63(0.10)  & 1.41(0.11) \\
 & 2048 & \hlc[light-gray]{0.19(0.01)}  & 0.25(0.05)  & 0.92(0.09)  & 1.41(0.09)  & 1.44(0.10) \\
\hline
\end{tabular}
\end{center}
\end{table}

\begin{table}[H]
\caption{$T_3$ noise with SNR=3: Ratio of average MSE of listed method to average MSE of VisuShrink with corresponding standard deviation ($\times10^{-3}$) of MSE of listed method}
\label{table:T3_3}
\begin{center}
\begin{tabular}{ccccccc}
\hline
Function & n & LD Block & Nason & CMWS & UH & EBayes \\ 
\hline
 & 512 & \hlc[light-gray]{0.43(0.03)}  & \hlc[light-gray]{0.42(0.08)}  & 0.95(0.35)  & 1.17(0.37)  & 1.34(0.39) \\
Blip  & 1024 & \hlc[light-gray]{0.34(0.03)}  & 0.39(0.06)  & 0.88(0.24)  & 1.09(0.25)  & 1.38(0.27) \\
 & 2048 & \hlc[light-gray]{0.26(0.02)}  & 0.33(0.14)  & 0.91(0.24)  & 1.08(0.24)  & 1.43(0.26) \\
\hline
 & 512 & 0.98(18.83)  & 0.88(19.81)  & 0.81(39.75)  & \hlc[light-gray]{0.66(41.32)}  & 1.12(42.93) \\
Blocks  & 1024 & \hlc[light-gray]{0.62(4.28)}  & \hlc[light-gray]{0.63(7.50)}  & 0.84(27.36)  & 0.76(28.71)  & 1.24(31.25) \\
 & 2048 & \hlc[light-gray]{0.44(2.55)}  & 0.49(15.57)  & 0.85(27.04)  & 0.82(27.80)  & 1.32(29.10) \\
\hline
 & 512 & 1.31(3.25)  & 1.58(4.55)  & \hlc[light-gray]{0.68(3.92)}  & 0.94(4.08)  & 1.01(4.25) \\
Bumps  & 1024 & 1.05(1.99)  & 1.61(4.26)  & \hlc[light-gray]{0.77(2.74)}  & 1.13(2.91)  & 1.21(3.08) \\
 & 2048 & \hlc[light-gray]{0.65(0.73)}  & 0.95(2.58)  & 0.84(2.73)  & 1.20(2.76)  & 1.33(2.89) \\
\hline
 & 512 & \hlc[light-gray]{0.19(0.02)}  & 0.24(0.05)  & 0.93(0.24)  & 1.30(0.24)  & 1.34(0.26) \\
Corner  & 1024 & \hlc[light-gray]{0.19(0.01)}  & 0.24(0.04)  & 0.92(0.16)  & 1.25(0.17)  & 1.41(0.18) \\
 & 2048 & \hlc[light-gray]{0.19(0.01)}  & 0.25(0.09)  & 0.92(0.16)  & 1.16(0.16)  & 1.44(0.17) \\
\hline
 & 512 & 0.74(0.23)  & \hlc[light-gray]{0.70(0.24)}  & \hlc[light-gray]{0.78(0.77)}  & 1.41(0.80)  & 1.23(0.82) \\
Doppler  & 1024 & \hlc[light-gray]{0.38(0.14)}  & 0.59(0.15)  & 0.84(0.52)  & 1.43(0.54)  & 1.34(0.60) \\
 & 2048 & \hlc[light-gray]{0.34(0.05)}  & 0.39(0.30)  & 0.88(0.52)  & 1.35(0.52)  & 1.42(0.55) \\
\hline
 & 512 & \hlc[light-gray]{0.20(7.14)}  & 0.25(15.53)  & 0.93(80.76)  & 1.37(84.62)  & 1.35(89.06) \\
Heavisine  & 1024 & \hlc[light-gray]{0.21(4.70)}  & 0.25(14.34)  & 0.92(54.36)  & 1.29(56.92)  & 1.40(63.30) \\
 & 2048 & \hlc[light-gray]{0.20(4.62)}  & 0.26(31.10)  & 0.92(54.45)  & 1.18(55.46)  & 1.44(58.46) \\
\hline
 & 512 & \hlc[light-gray]{0.86(377.77)}  & 0.92(634.93)  & \hlc[light-gray]{0.83(923.13)}  & 1.17(957.29)  & 1.26(1002.18) \\
Spikes  & 1024 & \hlc[light-gray]{0.43(142.21)}  & 0.48(184.45)  & 0.89(618.77)  & 1.27(655.00)  & 1.39(714.08) \\
 & 2048 & \hlc[light-gray]{0.21(57.57)}  & 0.27(352.48)  & 0.91(617.48)  & 1.22(619.83)  & 1.44(664.51) \\
\hline
 & 512 & \hlc[light-gray]{0.36(0.04)}  & \hlc[light-gray]{0.38(0.04)}  & 0.81(0.23)  & 1.40(0.24)  & 1.18(0.25) \\
Wave  & 1024 & \hlc[light-gray]{0.21(0.02)}  & 0.25(0.04)  & 0.92(0.15)  & 1.55(0.17)  & 1.41(0.18) \\
 & 2048 & \hlc[light-gray]{0.19(0.01)}  & 0.25(0.09)  & 0.92(0.15)  & 1.33(0.16)  & 1.44(0.17) \\
\hline
\end{tabular}
\end{center}
\end{table}

\begin{table}[H]
\caption{Lognormal noise with SNR=5: Ratio of average MSE of listed method to average MSE of VisuShrink with corresponding standard deviation ($\times10^{-3}$) of MSE of listed method} 
\label{table:Lognormal_5}
\begin{center}
\begin{tabular}{ccccccc}
\hline
Function & n & LD Block & Nason & CMWS & UH & EBayes \\ 
\hline
 & 512 & 0.37(0.04)  & \hlc[light-gray]{0.31(0.06)}  & 0.97(0.20)  & 1.24(0.20)  & 1.34(0.21) \\
Blip  & 1024 & \hlc[light-gray]{0.23(0.03)}  & 0.31(0.07)  & 0.96(0.19)  & 1.21(0.19)  & 1.35(0.19) \\
 & 2048 & \hlc[light-gray]{0.17(0.02)}  & 0.21(0.05)  & 0.96(0.13)  & 1.20(0.13)  & 1.39(0.14) \\
\hline
 & 512 & 1.23(18.94)  & 0.99(15.49)  & \hlc[light-gray]{0.90(22.09)}  & 0.97(22.70)  & 1.18(23.23) \\
Blocks  & 1024 & \hlc[light-gray]{0.58(3.23)}  & 0.62(8.11)  & 0.92(21.55)  & 1.03(21.72)  & 1.25(22.02) \\
 & 2048 & \hlc[light-gray]{0.39(2.36)}  & 0.44(4.91)  & 0.93(14.80)  & 1.09(15.28)  & 1.33(15.89) \\
\hline
 & 512 & 1.56(2.28)  & 1.97(3.47)  & \hlc[light-gray]{0.83(2.15)}  & 1.04(2.24)  & 1.13(2.27) \\
Bumps  & 1024 & \hlc[light-gray]{0.93(1.57)}  & 1.55(3.96)  & \hlc[light-gray]{0.90(2.18)}  & 1.15(2.17)  & 1.25(2.24) \\
 & 2048 & \hlc[light-gray]{0.58(0.72)}  & 0.85(1.79)  & 0.93(1.49)  & 1.21(1.51)  & 1.34(1.59) \\
\hline
 & 512 & \hlc[light-gray]{0.12(0.02)}  & 0.17(0.03)  & 0.97(0.13)  & 1.27(0.13)  & 1.33(0.14) \\
Corner  & 1024 & \hlc[light-gray]{0.11(0.01)}  & 0.15(0.05)  & 0.97(0.13)  & 1.24(0.13)  & 1.35(0.13) \\
 & 2048 & \hlc[light-gray]{0.11(0.01)}  & 0.14(0.03)  & 0.96(0.09)  & 1.22(0.09)  & 1.39(0.09) \\
\hline
 & 512 & \hlc[light-gray]{0.71(0.23)}  & \hlc[light-gray]{0.67(0.22)}  & 0.90(0.42)  & 1.32(0.43)  & 1.26(0.45) \\
Doppler  & 1024 & \hlc[light-gray]{0.25(0.11)}  & 0.49(0.20)  & 0.94(0.41)  & 1.28(0.41)  & 1.31(0.42) \\
 & 2048 & \hlc[light-gray]{0.26(0.04)}  & 0.29(0.09)  & 0.94(0.28)  & 1.27(0.29)  & 1.37(0.30) \\
\hline
 & 512 & \hlc[light-gray]{0.13(6.26)}  & 0.18(11.97)  & 0.97(44.70)  & 1.28(45.81)  & 1.32(47.74) \\
Heavisine  & 1024 & \hlc[light-gray]{0.12(4.34)}  & 0.17(16.17)  & 0.96(43.18)  & 1.24(43.51)  & 1.34(44.37) \\
 & 2048 & \hlc[light-gray]{0.11(2.87)}  & 0.15(9.76)  & 0.96(29.57)  & 1.23(30.44)  & 1.39(32.14) \\
\hline
 & 512 & \hlc[light-gray]{0.75(331.14)}  & \hlc[light-gray]{0.72(426.81)}  & 0.92(508.73)  & 1.17(516.66)  & 1.25(532.81) \\
Spikes  & 1024 & \hlc[light-gray]{0.25(110.09)}  & 0.36(197.79)  & 0.95(489.52)  & 1.21(492.82)  & 1.32(502.77) \\
 & 2048 & \hlc[light-gray]{0.13(34.40)}  & 0.16(110.58)  & 0.95(336.12)  & 1.23(342.25)  & 1.38(363.60) \\
\hline
 & 512 & \hlc[light-gray]{0.31(0.02)}  & 0.35(0.03)  & 0.89(0.13)  & 1.24(0.13)  & 1.19(0.14) \\
Wave  & 1024 & \hlc[light-gray]{0.11(0.01)}  & 0.16(0.05)  & 0.97(0.12)  & 1.33(0.12)  & 1.35(0.13) \\
 & 2048 & \hlc[light-gray]{0.11(0.01)}  & 0.14(0.03)  & 0.96(0.08)  & 1.29(0.09)  & 1.39(0.09) \\
\hline
\end{tabular}
\end{center}
\end{table}

\begin{table}[H]
\caption{Lognormal noise with SNR=3: Ratio of average MSE of listed method to average MSE of VisuShrink with corresponding standard deviation ($\times10^{-3}$) of MSE of listed method}
\label{table:Lognormal_3}
\begin{center}
\begin{tabular}{ccccccc}
\hline
Function & n & LD Block & Nason & CMWS & UH & EBayes \\ 
\hline
 & 512 & \hlc[light-gray]{0.28(0.07)}  & \hlc[light-gray]{0.29(0.09)}  & 0.97(0.33)  & 1.23(0.33)  & 1.34(0.35) \\
Blip  & 1024 & \hlc[light-gray]{0.20(0.04)}  & 0.25(0.12)  & 0.96(0.31)  & 1.19(0.31)  & 1.34(0.32) \\
 & 2048 & \hlc[light-gray]{0.15(0.03)}  & 0.19(0.07)  & 0.95(0.21)  & 1.19(0.22)  & 1.39(0.23) \\
\hline
 & 512 & \hlc[light-gray]{0.78(19.29)}  & \hlc[light-gray]{0.76(19.78)}  & 0.90(36.89)  & 0.98(37.80)  & 1.19(38.85) \\
Blocks  & 1024 & \hlc[light-gray]{0.39(3.92)}  & 0.44(13.64)  & 0.92(35.87)  & 1.04(36.24)  & 1.25(36.77) \\
 & 2048 & \hlc[light-gray]{0.28(2.98)}  & 0.32(8.18)  & 0.93(24.69)  & 1.09(25.39)  & 1.32(26.57) \\
\hline
 & 512 & 1.17(3.17)  & 1.44(4.22)  & \hlc[light-gray]{0.83(3.59)}  & 1.03(3.75)  & 1.13(3.79) \\
Bumps  & 1024 & \hlc[light-gray]{0.76(2.64)}  & 1.21(4.72)  & 0.91(3.64)  & 1.14(3.63)  & 1.25(3.72) \\
 & 2048 & \hlc[light-gray]{0.43(0.76)}  & 0.67(2.11)  & 0.93(2.47)  & 1.20(2.52)  & 1.34(2.65) \\
\hline
 & 512 & \hlc[light-gray]{0.12(0.03)}  & 0.16(0.06)  & 0.97(0.22)  & 1.26(0.22)  & 1.33(0.23) \\
Corner  & 1024 & \hlc[light-gray]{0.11(0.02)}  & 0.15(0.08)  & 0.97(0.21)  & 1.23(0.21)  & 1.35(0.22) \\
 & 2048 & \hlc[light-gray]{0.11(0.01)}  & 0.14(0.05)  & 0.96(0.14)  & 1.21(0.15)  & 1.39(0.16) \\
\hline
 & 512 & \hlc[light-gray]{0.54(0.20)}  & \hlc[light-gray]{0.53(0.26)}  & 0.90(0.70)  & 1.27(0.73)  & 1.25(0.75) \\
Doppler  & 1024 & \hlc[light-gray]{0.22(0.16)}  & 0.40(0.27)  & 0.94(0.68)  & 1.26(0.69)  & 1.32(0.70) \\
 & 2048 & \hlc[light-gray]{0.20(0.06)}  & 0.23(0.15)  & 0.94(0.47)  & 1.26(0.48)  & 1.37(0.51) \\
\hline
 & 512 & \hlc[light-gray]{0.12(9.90)}  & 0.17(19.91)  & 0.97(74.50)  & 1.28(76.21)  & 1.33(79.83) \\
Heavisine  & 1024 & \hlc[light-gray]{0.11(7.23)}  & 0.16(26.95)  & 0.97(72.00)  & 1.24(72.32)  & 1.35(74.13) \\
 & 2048 & \hlc[light-gray]{0.11(4.79)}  & 0.15(16.26)  & 0.96(49.30)  & 1.22(50.74)  & 1.39(53.60) \\
\hline
 & 512 & \hlc[light-gray]{0.60(439.61)}  & 0.67(529.10)  & 0.93(846.93)  & 1.17(870.70)  & 1.26(889.24) \\
Spikes  & 1024 & \hlc[light-gray]{0.24(152.73)}  & 0.30(313.31)  & 0.96(815.74)  & 1.21(822.55)  & 1.33(839.48) \\
 & 2048 & \hlc[light-gray]{0.12(61.61)}  & 0.15(184.31)  & 0.95(560.22)  & 1.22(572.79)  & 1.38(605.96) \\
\hline
 & 512 & \hlc[light-gray]{0.23(0.03)}  & 0.28(0.06)  & 0.91(0.21)  & 1.24(0.22)  & 1.22(0.23) \\
Wave  & 1024 & \hlc[light-gray]{0.11(0.02)}  & 0.16(0.08)  & 0.97(0.20)  & 1.30(0.21)  & 1.35(0.21) \\
 & 2048 & \hlc[light-gray]{0.11(0.01)}  & 0.14(0.05)  & 0.96(0.14)  & 1.27(0.14)  & 1.39(0.15) \\
\hline
\end{tabular}
\end{center}
\end{table}

We considered three different classifications for the results of the proposed method (LD Block). Wins are cases where LD Block has the lowest average MSE and this average MSE is significantly different than all other competitors in pairwise comparisons. Losses are cases where LD Block has a significantly different average MSE than the leader. Ties are cases where LD Block is not significantly different than the leader or LD Block is the leader but is not significantly different than all other methods. 

For $T_3$ noise and SNR=5, LD Block has 13 wins, 7 losses, and 3 ties. When SNR=3, LD Block has 16 wins, 4 losses, and 4 ties. Regardless of SNR, the proposed method loses only two of all the comparisons at the highest two sample sizes ($n = 2^{10}$ and $2^{11}$). The only cases in which the proposed method is significantly worse than the leading method is for smaller sample sizes and test functions Blocks, Bumps, Doppler, and Spikes.

For lognormal noise and SNR=5, the proposed method has 18 wins, 3 losses, and 3 ties. The only time a competitor beats the proposed method in terms of MSE is at the smallest sample size and test functions Blips, Blocks, and Bumps. When SNR=3, LD Block has 20 wins, 1 loss, and 3 ties.

LD Block, Nason, and CMWS show substantial improvement over VisuShrink, UH, and EBayes for both $T_3$ and lognormal noise, particularly for smoother functions. This improvement increases as SNR decreases, as shown in the tables. Clearly, outside of CMWS, traditional methods that assume normality fail drastically when given heavy-tailed or skewed errors.

\begin{figure}[H]
\begin{center}
\includegraphics[width=4in, height=3in, angle=0]{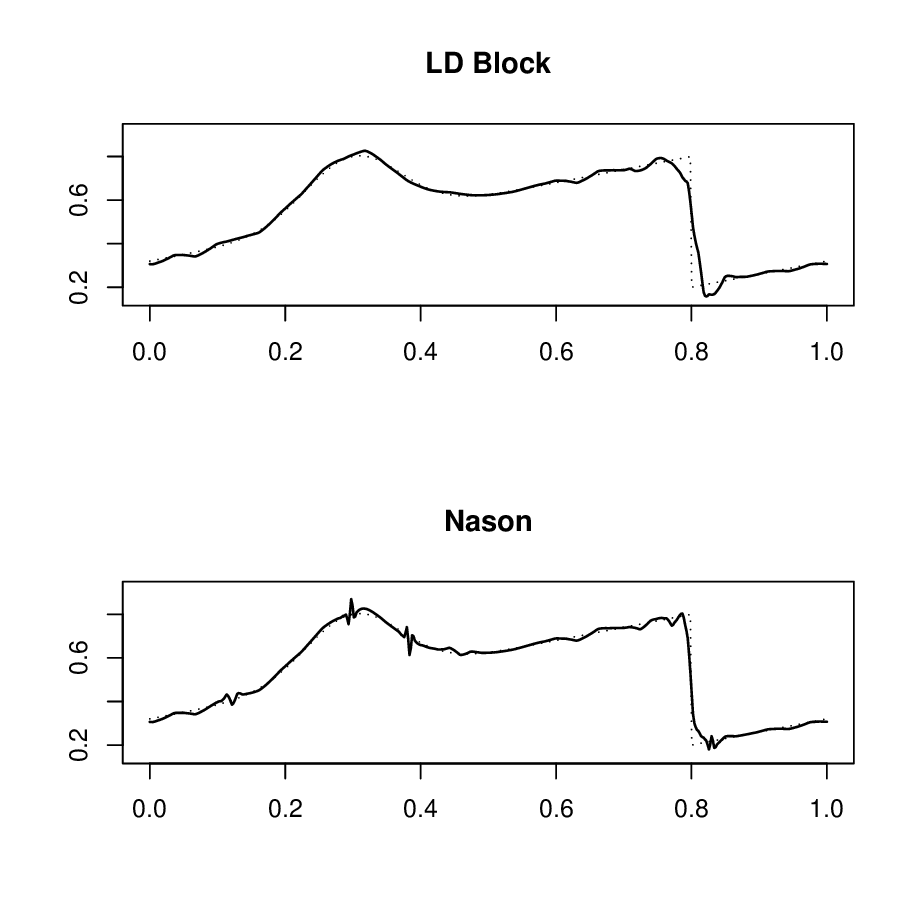}\\
\caption{Average reconstructions for our method and Nason's method using the Blip test function with $T_3$ noise, SNR=3, and n=512}
\label{figure:BlipReconT3}
\end{center}
\end{figure}

The proposed method also provides very visually appealing reconstructions for both smooth and spiky functions. Figure \ref{figure:BlipReconT3} shows the average reconstruction over 10 repetitions for the proposed method and Nason's method using the Blip test function, $T_3$ noise, SNR=3, and $n$=512. Both Nason's and the proposed method provide much smoother average reconstructions than does VisuShrink (see \cite{M1006} for VisuShrink reconstruction), however, it is clear that, in four or five places, Nason's method does not threshold a piece of data that it should have. This does not appear to happen at all for our level-dependent block method.

\begin{figure}[H]
\begin{center}
\includegraphics[width=4in, height=3in, angle=0]{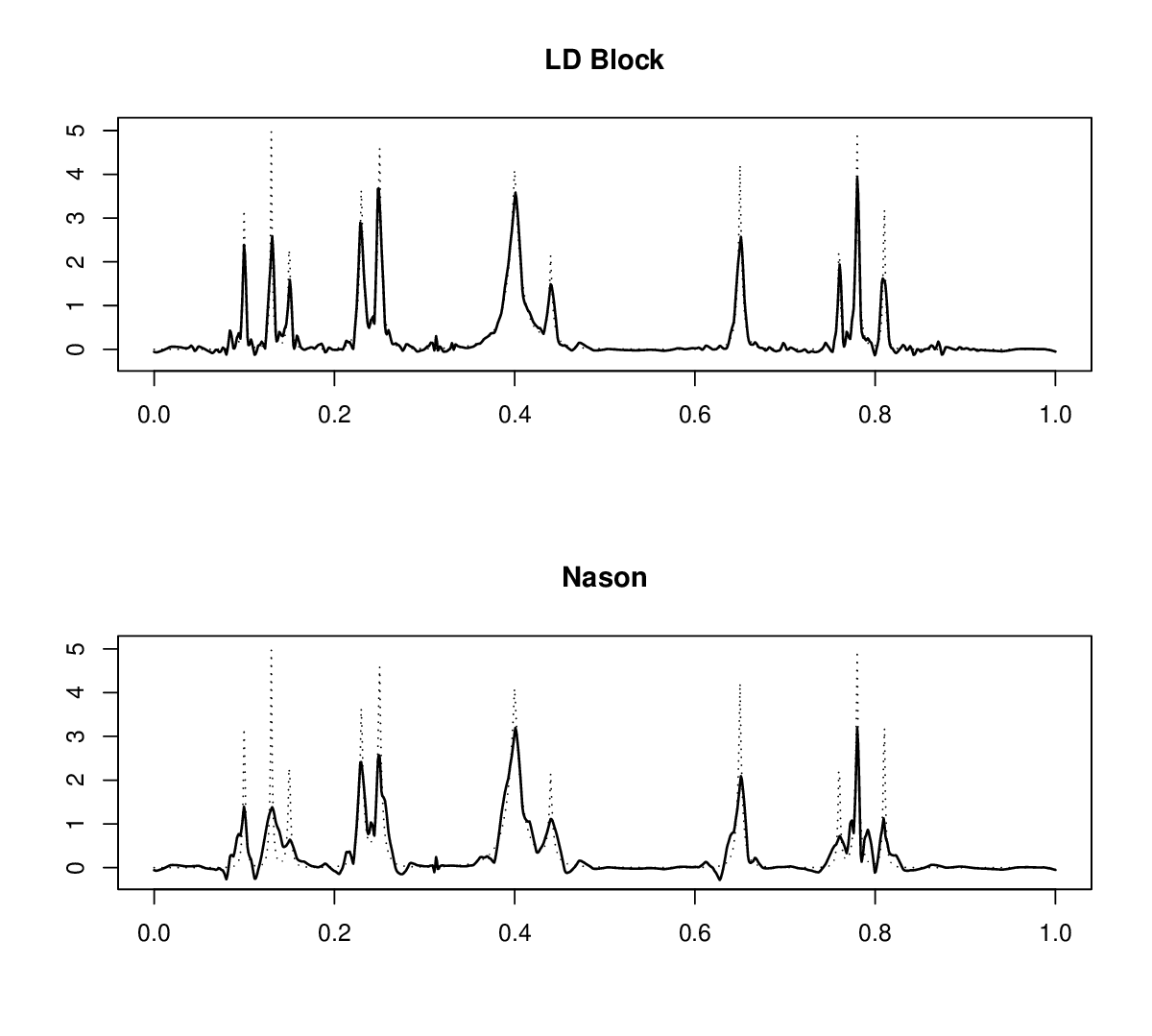}\\
\caption{Average reconstructions for our method and Nason's method using the Bumps test function with lognormal noise, SNR=3, and n=512}
\label{figure:BumpsReconLN}
\end{center}
\end{figure}

Figure \ref{figure:BumpsReconLN} shows the average reconstruction over 10 repetitions for the proposed method and Nason's method using the Bumps test function, lognormal noise, SNR=3, and $n$=1024. While VisuShrink does not smooth the function enough and the reconstruction has a lot of excess noise (see \cite{M1006}), Nason's reconstruction is overly smooth; it does not do a good job of picking up the jumps in the function and, in some cases, does not model them at all. The proposed method provides a ``happy medium'' visually. It models the spikes better than Nason's method, but without the excessive noise seen in VisuShrink's reconstruction. As a result, the MSE for our level-dependent block method is also the smallest of the three methods.

We tested the robustness of our method by comparing the performance of our estimator to current estimators with Gaussian errors. Tables \ref{table:Normal_5} and \ref{table:Normal_3} show these results. Our method is intended for use in situations when errors are thought to be non-normal, therefore we do not necessarily expect our method to be an improvement. For normal noise and SNR=5, LD Block has 24 losses. For normal noise and SNR=3, LD block has 1 tie and 23 losses. With Gaussian errors, CMWS appears to be the optimal method as it has the lowest average MSE in all but 7 of the 48 cases. CMWS outperforms LD Block in all 48 cases with normal noise and has an average MSE that is on average 1.84 times lower than that of LD Block. However, in the 96 cases with non-normal errors, LD Block outperforms CMWS 81 times and has an average MSE that is on average 4.36 times lower than that of CMWS. We see that we outperform CMWS when we are expected to and we do so by a much wider margin than when CMWS outperforms our method.

\begin{table}[H]
\caption{Normal noise with SNR=5: Ratio of MSE of listed method to MSE of VisuShrink with corresponding standard deviation ($\times10^{-3}$) of MSE of listed method}
\label{table:Normal_5}
\begin{center}
\begin{tabular}{ccccccc}
\hline
Function & n & LD Block & Nason & CMWS & UH & EBayes \\
\hline
 & 512 & 2.04(0.01)  & 1.07(0.02)  & \hlc[light-gray]{0.89(0.02)}  & 1.38(0.02)  & 1.18(0.03) \\
Blip  & 1024 & 1.20(0.02)  & 1.05(0.02)  & \hlc[light-gray]{0.68(0.01)}  & 0.90(0.01)  & 1.00(0.02) \\
 & 2048 & 1.14(0.01)  & 1.00(0.01)  & \hlc[light-gray]{0.81(0.01)}  & \hlc[light-gray]{0.81(0.01)}  & 1.02(0.01) \\
\hline
 & 512 & 1.81(18.87)  & 1.05(6.06)  & 0.64(2.71)  & \hlc[light-gray]{0.13(2.91)}  & 0.84(3.93) \\
Blocks  & 1024 & 1.40(1.88)  & 1.05(3.46)  & 0.68(1.82)  & \hlc[light-gray]{0.10(1.58)}  & 0.87(2.36) \\
 & 2048 & 1.11(1.59)  & 1.04(1.72)  & 0.65(0.97)  & \hlc[light-gray]{0.08(0.98)}  & 0.88(1.44) \\
\hline
 & 512 & 1.76(2.18)  & 1.94(2.08)  & \hlc[light-gray]{0.44(0.31)}  & 0.86(0.69)  & 0.71(0.50) \\
Bumps  & 1024 & 1.72(1.29)  & 2.14(1.21)  & \hlc[light-gray]{0.50(0.20)}  & 1.22(0.36)  & 0.84(0.34) \\
 & 2048 & 1.53(0.59)  & 0.94(0.22)  & \hlc[light-gray]{0.56(0.09)}  & 1.50(0.17)  & 0.87(0.16) \\
\hline
 & 512 & 0.98(0.01)  & 1.08(0.01)  & \hlc[light-gray]{0.86(0.01)}  & 2.67(0.02)  & 1.29(0.02) \\
Corner  & 1024 & 0.96(0.01)  & 1.07(0.01)  & \hlc[light-gray]{0.89(0.01)}  & 2.03(0.01)  & 1.13(0.01) \\
 & 2048 & 0.97(0.00)  & 1.03(0.01)  & \hlc[light-gray]{0.87(0.00)}  & 1.39(0.01)  & 1.06(0.01) \\
\hline
 & 512 & 1.54(0.19)  & 0.94(0.06)  & \hlc[light-gray]{0.46(0.04)}  & 2.01(0.10)  & 0.95(0.09) \\
Doppler  & 1024 & 0.83(0.06)  & 0.95(0.04)  & \hlc[light-gray]{0.57(0.02)}  & 2.38(0.05)  & 0.96(0.05) \\
 & 2048 & 1.39(0.02)  & 0.99(0.02)  & \hlc[light-gray]{0.67(0.02)}  & 2.42(0.03)  & 0.98(0.02) \\
\hline
 & 512 & 0.94(3.48)  & 1.06(4.34)  & \hlc[light-gray]{0.90(3.13)}  & 2.89(5.60)  & 1.25(7.73) \\
Heavisine  & 1024 & 0.96(1.99)  & 1.06(2.86)  & \hlc[light-gray]{0.87(1.97)}  & 2.29(3.35)  & 1.12(3.40) \\
 & 2048 & 0.97(1.46)  & 1.03(1.78)  & \hlc[light-gray]{0.87(1.37)}  & 1.61(2.12)  & 1.06(2.01) \\
\hline
 & 512 & 2.04(284.61)  & 0.95(62.18)  & \hlc[light-gray]{0.52(46.82)}  & 1.40(82.13)  & 0.98(107.79) \\
Spikes  & 1024 & 1.41(87.44)  & 1.01(38.48)  & \hlc[light-gray]{0.68(23.64)}  & 1.94(42.91)  & 1.04(45.91) \\
 & 2048 & 0.98(17.52)  & 1.03(20.41)  & \hlc[light-gray]{0.79(16.05)}  & 1.97(25.73)  & 1.06(21.71) \\
\hline
 & 512 & 0.97(0.01)  & 1.01(0.01)  & \hlc[light-gray]{0.45(0.01)}  & 2.18(0.02)  & 0.82(0.02) \\
Wave  & 1024 & 1.11(0.01)  & 1.06(0.01)  & \hlc[light-gray]{0.87(0.01)}  & 4.56(0.01)  & 1.12(0.01) \\
 & 2048 & 0.97(0.00)  & 1.03(0.01)  & \hlc[light-gray]{0.87(0.00)}  & 2.99(0.01)  & 1.06(0.01) \\
\hline
\end{tabular}
\end{center}
\end{table}

\begin{table}[H]
\caption{Normal noise with SNR=3: Ratio of MSE of listed method to MSE of VisuShrink with corresponding standard deviation ($\times10^{-3}$) of MSE of listed method}
\label{table:Normal_3}
\begin{center}
\begin{tabular}{ccccccc}
\hline
Function & n & LD Block & Nason & CMWS & UH & EBayes \\ 
\hline
 & 512 & 1.47(0.02)  & 1.00(0.03)  & \hlc[light-gray]{0.87(0.03)}  & 1.09(0.03)  & 1.13(0.06) \\
Blip  & 1024 & 1.05(0.02)  & 1.01(0.02)  & \hlc[light-gray]{0.68(0.02)}  & 0.75(0.02)  & 0.99(0.03) \\
 & 2048 & 1.11(0.02)  & 1.01(0.01)  & 0.84(0.01)  & \hlc[light-gray]{0.71(0.01)}  & 1.03(0.01) \\
\hline
 & 512 & 1.44(18.82)  & 0.96(5.86)  & 0.67(4.00)  & \hlc[light-gray]{0.22(5.56)}  & 0.90(6.58) \\
Blocks  & 1024 & 1.15(2.22)  & 0.99(3.47)  & 0.75(2.34)  & \hlc[light-gray]{0.17(2.71)}  & 0.91(3.63) \\
 & 2048 & 1.02(1.50)  & 0.99(1.60)  & 0.73(1.42)  & \hlc[light-gray]{0.12(1.52)}  & 0.96(1.94) \\
\hline
 & 512 & 1.54(2.79)  & 1.48(2.30)  & \hlc[light-gray]{0.51(0.54)}  & 0.96(1.05)  & 0.79(0.78) \\
Bumps  & 1024 & 1.43(1.62)  & 1.43(1.55)  & \hlc[light-gray]{0.52(0.34)}  & 1.09(0.57)  & 0.81(0.52) \\
 & 2048 & 1.28(0.55)  & 0.94(0.31)  & \hlc[light-gray]{0.58(0.15)}  & 1.32(0.26)  & 0.86(0.26) \\
\hline
 & 512 & 0.98(0.02)  & 1.08(0.02)  & \hlc[light-gray]{0.88(0.02)}  & 2.34(0.03)  & 1.30(0.04) \\
Corner  & 1024 & 0.96(0.01)  & 1.07(0.01)  & \hlc[light-gray]{0.89(0.01)}  & 1.69(0.01)  & 1.13(0.02) \\
 & 2048 & 0.97(0.01)  & 1.03(0.01)  & \hlc[light-gray]{0.87(0.01)}  & 1.18(0.01)  & 1.06(0.01) \\
\hline
 & 512 & 1.50(0.14)  & 0.99(0.08)  & \hlc[light-gray]{0.55(0.07)}  & 2.03(0.14)  & 1.05(0.14) \\
Doppler  & 1024 & 0.83(0.09)  & 0.97(0.06)  & \hlc[light-gray]{0.61(0.04)}  & 2.09(0.08)  & 0.97(0.07) \\
 & 2048 & 1.12(0.03)  & 1.00(0.03)  & \hlc[light-gray]{0.70(0.02)}  & 2.08(0.04)  & 0.97(0.03) \\
\hline
 & 512 & \hlc[light-gray]{0.93(5.31)}  & 1.07(7.25)  & \hlc[light-gray]{0.90(5.20)}  & 2.48(9.00)  & 1.27(13.05) \\
Heavisine  & 1024 & 0.96(3.32)  & 1.06(4.75)  & \hlc[light-gray]{0.88(3.29)}  & 1.95(4.91)  & 1.13(5.83) \\
 & 2048 & 0.97(2.43)  & 1.03(2.96)  & \hlc[light-gray]{0.87(2.28)}  & 1.43(3.35)  & 1.06(3.36) \\
\hline
 & 512 & 1.89(365.21)  & 0.97(144.77)  & \hlc[light-gray]{0.55(71.82)}  & 1.31(132.43)  & 0.99(165.34) \\
Spikes  & 1024 & 1.32(95.74)  & 0.99(61.06)  & \hlc[light-gray]{0.68(39.27)}  & 1.67(67.18)  & 0.99(76.18) \\
 & 2048 & 0.98(30.43)  & 1.03(34.12)  & \hlc[light-gray]{0.83(26.45)}  & 1.74(37.28)  & 1.06(35.89) \\
\hline
 & 512 & 0.95(0.01)  & 1.03(0.02)  & \hlc[light-gray]{0.54(0.02)}  & 2.29(0.03)  & 1.10(0.03) \\
Wave  & 1024 & 1.08(0.02)  & 1.06(0.01)  & \hlc[light-gray]{0.87(0.01)}  & 3.89(0.02)  & 1.13(0.02) \\
 & 2048 & 0.97(0.01)  & 1.03(0.01)  & \hlc[light-gray]{0.87(0.01)}  & 2.60(0.01)  & 1.06(0.01) \\
\hline
\end{tabular}
\end{center}
\end{table}

We want to show that our comparisons are valid despite the selection of a different wavelet basis for two of the competitors (CMWS and UH). CMWS requires a complex-valued wavelet basis which cannot be applied to any of the other methods. However, the UH method selects an optimal Haar-like basis as its filter, so we used a Haar basis across the other methods to standardize comparisons. The results of these comparisons are shown for normal and non-normal noise in Tables  \ref{table:haar} and \ref{table:haar2}, respectively. In the presence of normal noise, the standardization of the wavelet basis does not make UH the best thresholding technique as it only won or tied 7 of the 24 cases. This is fewer than the number of wins or ties by LD Block and EBayes, which had 8 and 9 such cases, respectively. In the presence of non-normal noise, which UH is not meant for, the gap between LD Block and UH is even greater as LD Block wins 18 of 24 cases while UH wins only 3. We acknowledge the fact that UH outperforms our method by a large factor when estimating the Blocks function in the presence of normal noise and attribute this to the fact that UH is designed to use a Haar basis, which is suited for estimating piecewise linear functions such as Blocks.

\pagebreak

\begin{table}[H]
\caption{Haar basis with normal noise and SNR=5: Ratio of MSE of listed method to MSE of VisuShrink with corresponding standard deviation ($\times10^{-3}$) of MSE of listed method*}
\label{table:haar}
\begin{center}
\begin{tabular}{cccccc}
\hline
Function & n & LD Block & Nason & UH & EBayes \\ 
\hline
 & 512 & 2.58(0.10)  & 1.39(0.01)  & 1.24(0.02)  & 1.20(0.03) \\
Blip  & 1024 & 2.06(0.01)  & 1.37(0.02)  & 1.16(0.01)  & 1.06(0.02) \\
 & 2048 & 1.86(0.02)  & 1.02(0.01)  & \hlc[light-gray]{0.96(0.01)}  & 1.06(0.01) \\
\hline
 & 512 & 2.53(9.13)  & 1.56(10.09)  & \hlc[light-gray]{0.16(2.91)}  & 0.84(4.43) \\
Blocks  & 1024 & 2.30(10.93)  & 1.91(8.04)  & \hlc[light-gray]{0.14(1.58)}  & 0.94(2.61) \\
 & 2048 & 1.63(3.94)  & 1.58(2.73)  & \hlc[light-gray]{0.10(0.98)}  & 0.87(1.45) \\
\hline
 & 512 & 2.56(2.62)  & 2.62(2.71)  & 0.89(0.69)  & \hlc[light-gray]{0.73(0.48)} \\
Bumps  & 1024 & 1.44(1.52)  & 2.07(1.55)  & 1.06(0.36)  & \hlc[light-gray]{0.91(0.31)} \\
 & 2048 & 1.10(0.32)  & 1.02(0.30)  & 0.95(0.17)  & \hlc[light-gray]{0.85(0.20)} \\
\hline
 & 512 & \hlc[light-gray]{0.97(0.01)}  & 1.07(0.01)  & 1.63(0.02)  & 1.16(0.02) \\
Corner  & 1024 & \hlc[light-gray]{0.98(0.01)}  & 1.05(0.01)  & 1.64(0.01)  & 1.10(0.01) \\
 & 2048 & \hlc[light-gray]{0.98(0.00)}  & 1.03(0.01)  & 1.32(0.01)  & 1.06(0.01) \\
\hline
 & 512 & 1.14(0.27)  & 0.95(0.09)  & 1.11(0.10)  & \hlc[light-gray]{0.82(0.07)} \\
Doppler  & 1024 & 1.06(0.10)  & 0.96(0.05)  & 1.24(0.05)  & \hlc[light-gray]{0.87(0.05)} \\
 & 2048 & \hlc[light-gray]{0.87(0.04)}  & 0.97(0.02)  & 1.32(0.03)  & 0.92(0.02) \\
\hline
 & 512 & \hlc[light-gray]{0.98(3.37)}  & 1.04(4.82)  & 1.60(5.60)  & 1.10(6.27) \\
Heavisine  & 1024 & \hlc[light-gray]{1.01(2.76)}  & 1.05(2.93)  & 1.94(3.35)  & 1.11(3.36) \\
 & 2048 & \hlc[light-gray]{0.98(1.45)}  & 1.02(1.74)  & 1.50(2.12)  & 1.04(2.17) \\
\hline
 & 512 & 1.43(631.83)  & \hlc[light-gray]{0.96(85.78)}  & \hlc[light-gray]{0.94(82.13)}  & 0.97(98.93) \\
Spikes  & 1024 & 1.50(160.68)  & 0.96(45.71)  & \hlc[light-gray]{0.93(42.91)}  & \hlc[light-gray]{0.93(49.32)} \\
 & 2048 & 1.11(93.28)  & \hlc[light-gray]{0.98(23.47)}  & \hlc[light-gray]{0.96(25.73)}  & \hlc[light-gray]{0.97(29.33)} \\
\hline
 & 512 & 1.19(0.01)  & 0.96(0.02)  & 1.08(0.02)  & \hlc[light-gray]{0.67(0.02)} \\
Wave  & 1024 & \hlc[light-gray]{0.98(0.01)}  & 1.03(0.01)  & 1.61(0.01)  & \hlc[light-gray]{0.98(0.01)} \\
 & 2048 & 1.09(0.01)  & 1.02(0.00)  & 2.00(0.01)  & 1.07(0.01) \\
\hline
\end{tabular}
\end{center}
* Rows without a highlighted average MSE are cases in which VisuShrink had the lowest average MSE
\end{table}

\begin{table}[H]
\caption{Haar basis with $T_3$ noise and SNR=5: Ratio of MSE of listed method to MSE of VisuShrink with corresponding standard deviation ($\times10^{-3}$) of MSE of listed method*}
\label{table:haar2}
\begin{center}
\begin{tabular}{cccccc}
\hline
Function & n & LD Block & Nason & UH & Ebayes \\ 
\hline
 & 512 & 0.54(0.02)  & \hlc[light-gray]{0.45(0.06)}  & 1.10(0.22)  & 1.36(0.23) \\
Blip  & 1024 & \hlc[light-gray]{0.41(0.02)}  & \hlc[light-gray]{0.43(0.03)}  & 1.10(0.15)  & 1.43(0.16) \\
 & 2048 & \hlc[light-gray]{0.28(0.02)}  & 0.31(0.03)  & 1.06(0.15)  & 1.46(0.15) \\
\hline
 & 512 & 1.71(19.32)  & 1.44(16.60)  & \hlc[light-gray]{0.72(25.04)}  & 1.24(26.21) \\
Blocks  & 1024 & 1.01(2.99)  & 1.35(13.16)  & \hlc[light-gray]{0.82(17.11)}  & 1.38(18.40) \\
 & 2048 & \hlc[light-gray]{0.62(2.42)}  & 0.85(5.28)  & 0.86(16.71)  & 1.39(17.48) \\
\hline
 & 512 & 1.82(2.04)  & 2.87(3.59)  & \hlc[light-gray]{0.99(2.45)}  & 1.09(2.51) \\
Bumps  & 1024 & 1.19(1.45)  & 2.23(3.71)  & 1.06(1.76)  & 1.21(1.83) \\
 & 2048 & \hlc[light-gray]{0.67(0.72)}  & 1.41(2.57)  & 1.02(1.67)  & 1.23(1.72) \\
\hline
 & 512 & \hlc[light-gray]{0.17(0.01)}  & 0.29(0.01)  & 1.12(0.14)  & 1.31(0.16) \\
Corner  & 1024 & \hlc[light-gray]{0.17(0.01)}  & 0.22(0.01)  & 1.15(0.10)  & 1.42(0.11) \\
 & 2048 & \hlc[light-gray]{0.18(0.01)}  & 0.19(0.01)  & 1.11(0.10)  & 1.45(0.10) \\
\hline
 & 512 & \hlc[light-gray]{0.64(0.24)}  & 0.95(0.20)  & 1.05(0.49)  & 1.01(0.48) \\
Doppler  & 1024 & \hlc[light-gray]{0.30(0.11)}  & 0.79(0.08)  & 1.11(0.32)  & 1.18(0.35) \\
 & 2048 & \hlc[light-gray]{0.34(0.03)}  & 0.58(0.11)  & 1.12(0.31)  & 1.30(0.33) \\
\hline
 & 512 & \hlc[light-gray]{0.17(4.28)}  & 0.34(5.01)  & 1.16(50.53)  & 1.28(52.70) \\
Heavisine  & 1024 & \hlc[light-gray]{0.20(2.82)}  & 0.24(3.05)  & 1.20(34.44)  & 1.41(37.15) \\
 & 2048 & \hlc[light-gray]{0.19(2.77)}  & 0.21(3.14)  & 1.13(33.48)  & 1.44(35.22) \\
\hline
 & 512 & \hlc[light-gray]{0.90(345.62)}  & 1.03(754.27)  & 1.00(571.63)  & 1.18(596.81) \\
Spikes  & 1024 & \hlc[light-gray]{0.36(111.53)}  & 0.83(348.20)  & 1.01(386.95)  & 1.29(411.52) \\
 & 2048 & \hlc[light-gray]{0.18(37.16)}  & 0.58(100.40)  & 1.02(373.10)  & 1.37(397.86) \\
\hline
 & 512 & \hlc[light-gray]{0.34(0.03)}  & 0.77(0.03)  & 1.03(0.15)  & 1.02(0.15) \\
Wave  & 1024 & \hlc[light-gray]{0.15(0.01)}  & 0.41(0.01)  & 1.13(0.10)  & 1.22(0.10) \\
 & 2048 & \hlc[light-gray]{0.16(0.01)}  & 0.25(0.01)  & 1.22(0.09)  & 1.43(0.10) \\
\hline
\end{tabular}
\end{center}
* Rows without a highlighted average MSE are cases in which VisuShrink had the lowest average MSE
\end{table}

\begin{figure}[H]
\begin{center}
\includegraphics[width=4in, height=3in, angle=0]{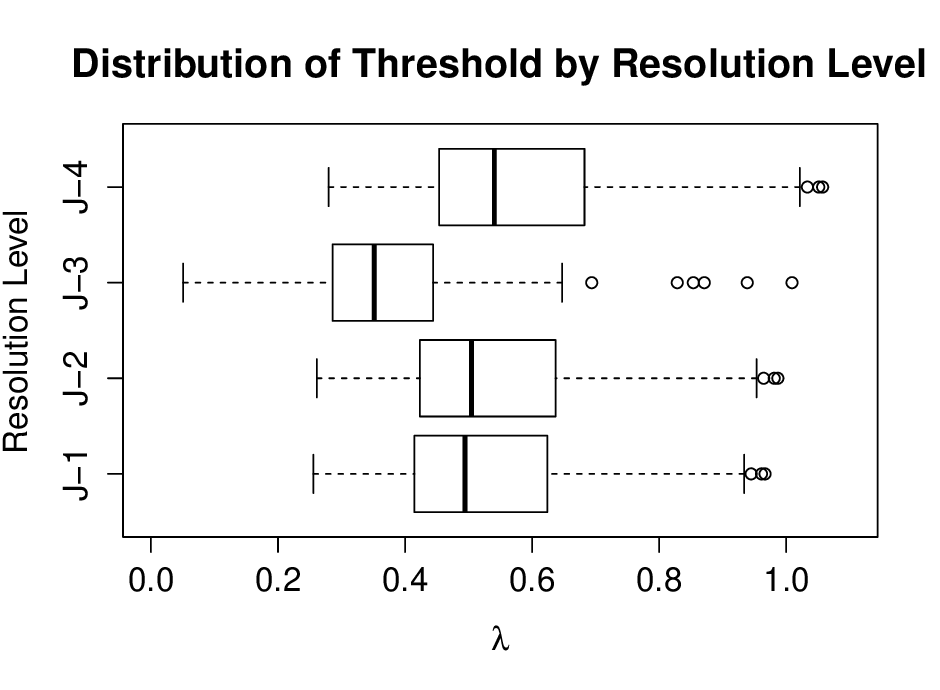}\\
\caption{Distributions of $\lambda$ by resolution level for 100 simulations of the Blip function with $T_3$ noise and SNR=5.}
\label{figure:LambdaVarianceComp}
\end{center}
\end{figure}

\begin{figure}[H]
\begin{center}
\includegraphics[width=4in, height=3in, angle=0]{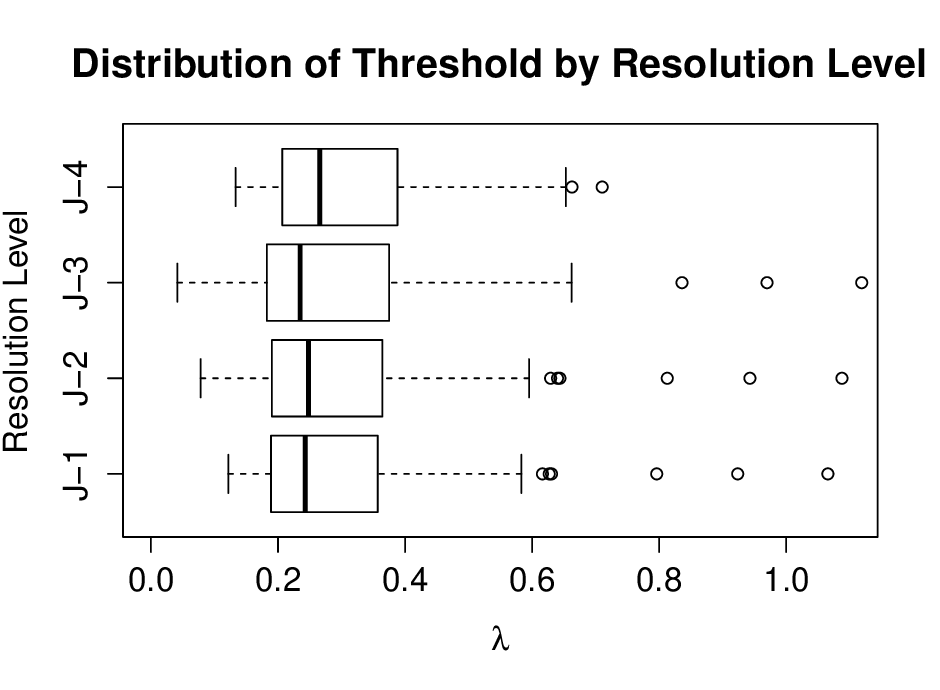}\\
\caption{Distributions of $\lambda$ by resolution level for 100 simulations of the Wave function with $T_3$ noise and SNR=5.}
\label{figure:LambdaVarianceComp2}
\end{center}
\end{figure}

To explore the distribution of $\lambda$ by resolution level, we studied the distribution of 100 threshold values for each of four resolution levels when estimating the Blip function with $T_3$ noise and SNR=5. While the variance of $\lambda$ was greater for coarse scales with fewer wavelet coefficients, the difference between the variance of $\lambda$ at the highest resolution level and at the lowest resolution level was not significant. A visual comparison of the distributions is given in Figure \ref{figure:LambdaVarianceComp} and shows comparable distributions of the threshold values across all resolution levels except $J-3$. This can primarily be attributed to the choice of function, Blip, as evidenced in in Figure \ref{figure:LambdaVarianceComp2} where we observe the distributions of threshold values under the same conditions as before but for the Wave function and note comparable distributions across all levels. The similar pattern of outliers shown in Figures \ref{figure:LambdaVarianceComp} and \ref{figure:LambdaVarianceComp2} indicates highly correlations between the values of $\lambda$ for different resolution levels. As shown in Table \ref{table:cormatrix}, the Wave simulation described above yields high, positive correlations between threshold values, particularly for the three highest resolution levels. 
 
\begin{table}[H]
\caption{Correlations between $\lambda$ values from different resolution levels for 100 simulations of the Wave function with $T_3$ noise and SNR=5.}
\label{table:cormatrix}
\begin{center}
\begin{tabular}{c|cccc}
Resolution Level & $J-1$ & $J-2$ & $J-3$ & $J-4$ \\
\hline
$J-1$ & 1.00 \\
$J-2$ & 1.00 & 1.00 \\
$J-3$ & 0.99 & 0.99 & 1.00 \\
$J-4$ & 0.63 & 0.63 & 0.63 & 1.00

\end{tabular}
\end{center}
\end{table}

\subsection{Examples}

Respiratory Inductance Plethysmography (IP) is a method of evaluating pulmonary ventilation by measuring the variation in size of the chest and abdominal wall. If calibrated properly, the output voltage of the inductance plethysmograph is proportional to the change in volume of the body port under evaluation, in this case the lungs. 

\begin{figure}[H]
\begin{center}
\includegraphics[width=4in, height=3in, angle=0]{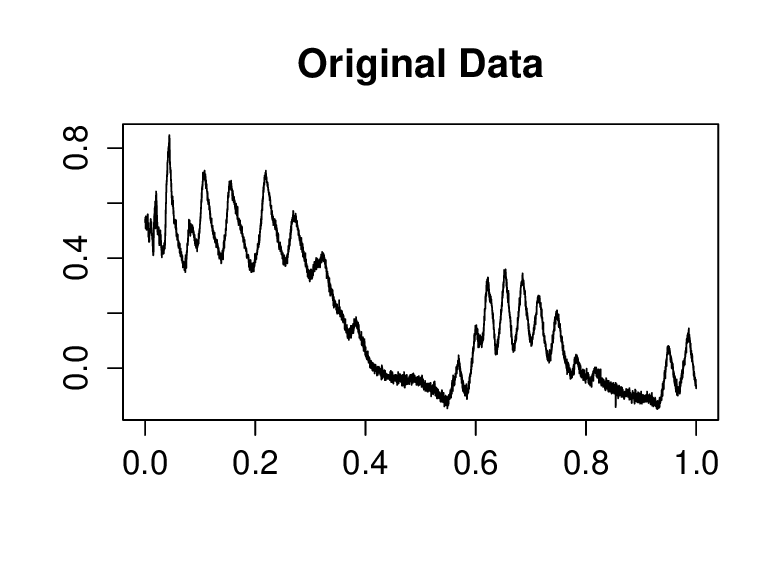}\\
\caption{Inductance Plethysmography Data}
\label{figure:IPDOrig}
\end{center}
\end{figure}

\begin{figure}[H]
\begin{center}
\includegraphics[width=6.5in, height=9.5in, angle=0]{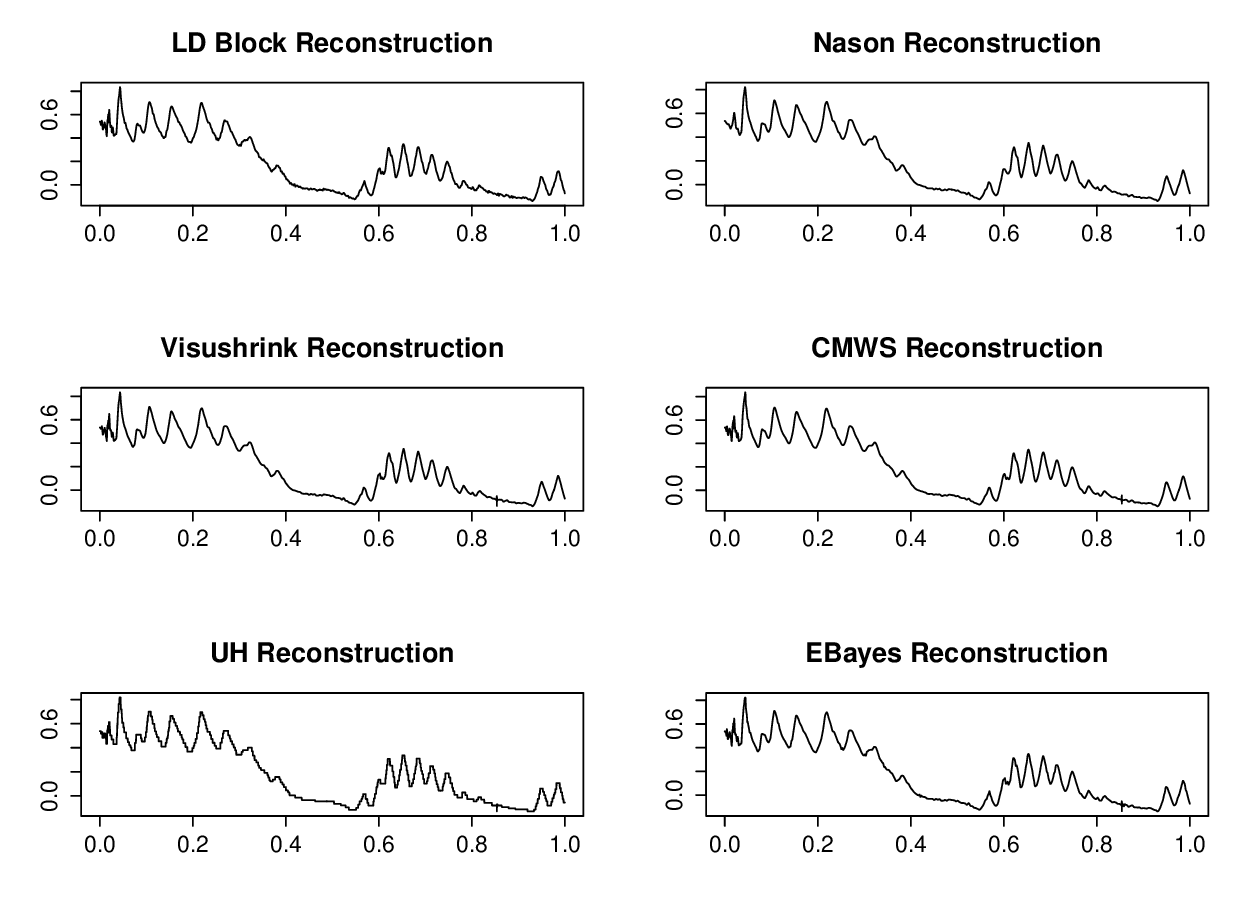}\\
\caption{Inductance Plethysmography Data Reconstruction}
\label{figure:IPDRecon}
\end{center}
\end{figure}

In \cite{nason2}, data from approximately 80 s of plehysmogorph recordings made by the Department of Anaesthesia at the Bristol Royal Infirmary were used to compare various thresholding methods. Figure \ref{figure:IPDOrig} shows the original IP data consisting of 4096 data points and Figure \ref{figure:IPDRecon} shows the reconstructions of all 6 thresholding methods in our comparisons. In order to avoid the introduction of artificial singularities at the boundaries of our reconstruction (unlike our simulated data, our actual data is not periodic), we extended the original data at both the beginning and end by reflecting the data about the boundaries. For computation, the data was extended to the next dyadic, but only the reconstruction of the observed portion of the data is pictured. In the figures, the two main sets of regular oscillations correspond to normal breathing, while the disturbed behavior in the middle of the profile corresponds to the patient vomiting. 

\pagebreak

With the exception of UH, all reconstructions are visually similar. The Nason reconstruction appears to be the most appealing as LD Block does not threshold some high-resolution coefficients that may come as a result of noise at the beginning of the data. Our method retains 12.7\% of the coefficients and sets the rest to zero while Nason's method retains 6.3\% of the coefficients. The threshold values for our LD Block method (with the lowest resolution level set to $J-4$) were $\lambda_{J-1} = 0.20$, $\lambda_{J-2} = 0.46$, $\lambda_{J-3} = 0.01$, and $\lambda_{J-4} = 0.00$. The threshold value for Nason's method was $\lambda = 0.44$. Since block thresholding thresholds the sum of squared coefficients in any block, in order to meaningfully compare the thresholds we must divide each LD Block threshold by the block size (L=8) then take the square root. This gives us $\lambda_{J-1}^* = 0.16$, $\lambda_{J-2}^* = 0.24$, $\lambda_{J-3}^* = 0.04$, and $\lambda_{J-4}^* = 0.00$. The reconstructions appear essentially the same, but since our threshold values are all smaller than Nason's threshold value, we do end up with slightly more noise at the left extreme of the profile. We also note that with a threshold value of zero at the lowest resolution level, our method did not threshold any of the coefficients in that level. 

\begin{figure}[H]
\begin{center}
\includegraphics[width=4in, height=3in, angle=0]{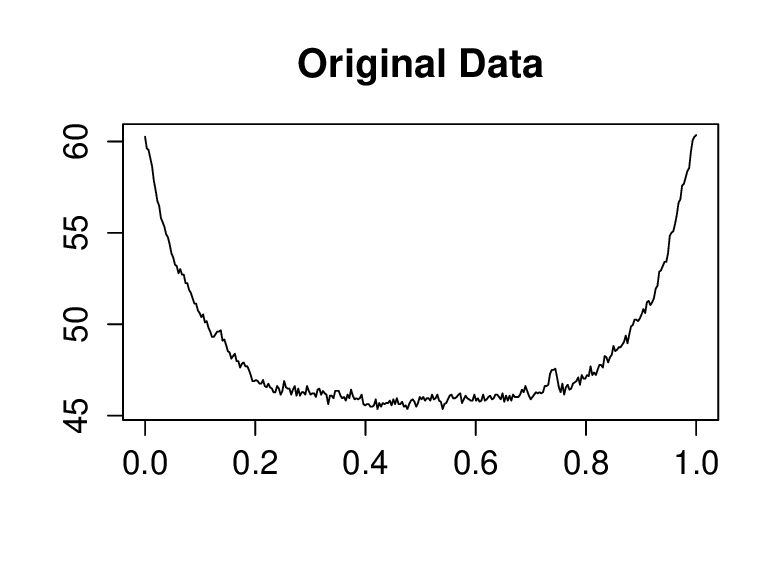}\\
\caption{Vertical Density Profile Data}
\label{figure:VDPOrig}
\end{center}
\end{figure}

\begin{figure}[H]
\begin{center}
\includegraphics[width=6.5in, height=9.5in, angle=0]{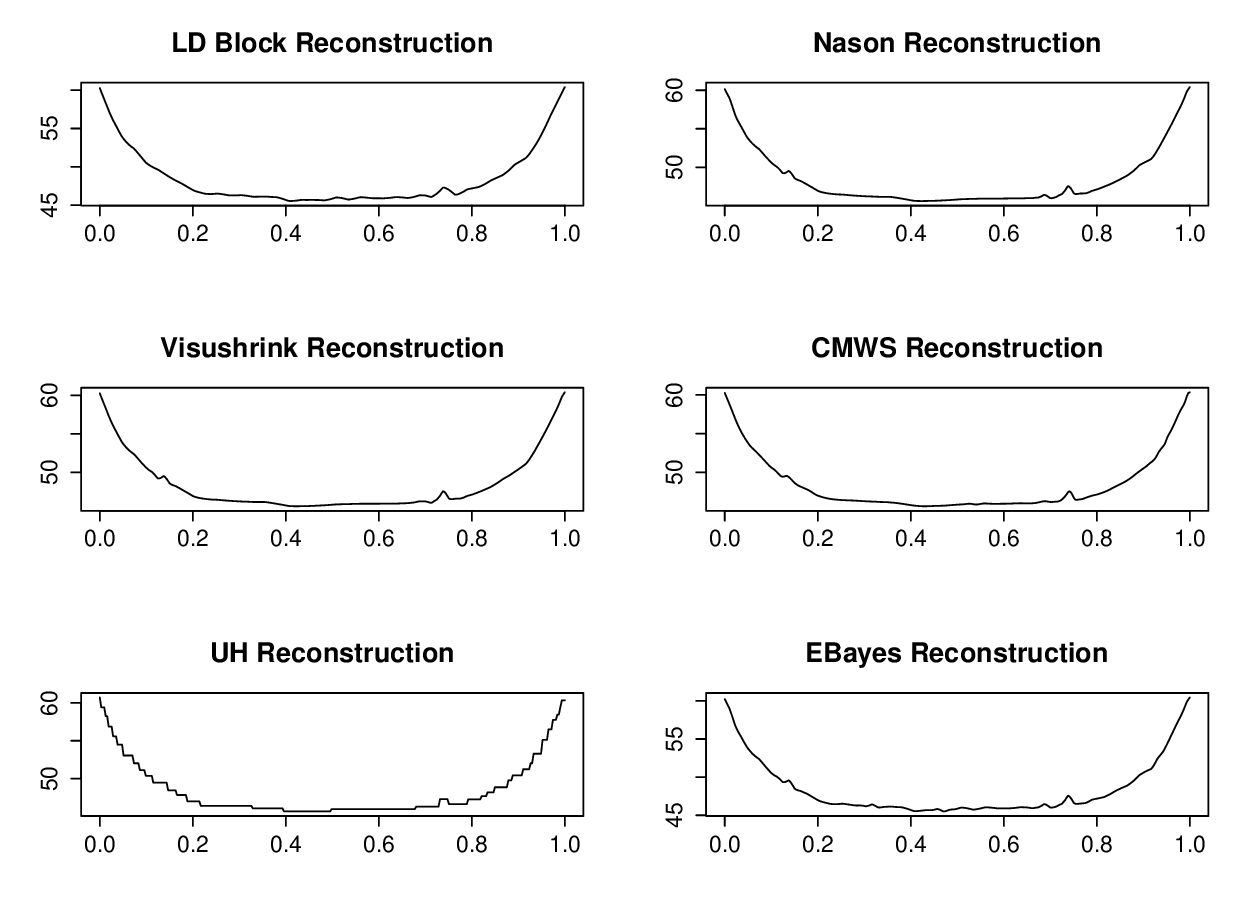}\\
\caption{Vertical Density Profile Data Reconstruction}
\label{figure:VDPRecon}
\end{center}
\end{figure}

In another relevant example with real data, we considered the vertical density profile data from \cite{Walker00comparingcurves}. As a measure of quality control in the manufacturing of particle board, the density of a particle board is measured at fixed vertical depths between the two faces of the board. The density measurements for a given board comprise its \textit{vertical density profile} (VDP). Figure \ref{figure:VDPOrig} shows the original VDP data of a single board which consisted of 314 observations and Figure \ref{figure:VDPRecon} shows the reconstructions of the VDP for each considered method. Similar to our handling of the non-periodic data in our first example, we addressed potential boundary issues by reflecting the data about both boundaries to extend original data to the next dyadic. 

\pagebreak

Again, with the exception of the more rigid reconstruction provided by the UH basis, reconstructions given in this example appear to be even more visually similar than in the last. There does appear to be a small peak around a vertical depth of 0.15 that our method thresholds and the other methods do not. Despite this observation, our method appears to retain a similar amount of detail throughout its reconstruction as it retains 12.5\% of coefficients while Nason's method retains only 10.6\% of coefficients. The universal threshold value for Nason's method was $\lambda = 0.46$. The thresholds from our method, after the same standardization for comparison described in our previous example, were $\lambda_{J-1}^* = 0.72$, $\lambda_{J-2}^* = 0.73$, $\lambda_{J-3}^* = 0.57$, and $\lambda_{J-4}^* = 0.25$.

\section{Discussion and Conclusion}

Our goal in developing the proposed method was to provide a completely non-parametric way to threshold wavelet coefficients when the distribution of the noise is unknown. Simulation results show that our level-dependent block thresholding method is an improvement over the existing cross-validated method as well as traditional methods for various types of non-Gaussian noise.

Other combinations of cross-validation and block thresholding methods were also considered. Generally speaking, global block thresholding (Block) was approximately equivalent to Nason's method in terms of MSE, while the term-by-term level-dependent version of Nason's method (LD) was an improvement over Nason's global method (Nason) but not as good as the proposed level-dependent block method (LD Block). A version of the proposed method that was block-dependent as opposed to level-dependent (BD Block) was considered as well, but it did not perform very well. Since this method allows for a different threshold value to be chosen for every individual block, it was a very local analysis and was likely trying to model too much of the noise. It was also very computationally expensive. (See \cite{M1006} for full results). Thus the  general MSE trend for non-normal noise was:

\begin{center}
LD Block $<$ LD $<$ Nason $\approx$ Block $<$ BD Block $<$ VisuShrink.
\end{center}

An alternate cross-validation method was considered, in which we randomly chose (without replacement) half of the noisy data on which to perform the DWT and thresholding, and then compared that reconstruction to the other half of the data. However this method did not perform quite as well, and was significantly slower than the even-odd method, since numerous repetitions of the random selection and cross-validation were needed to ensure valid reconstructions. 

\begin{figure}[H]
\begin{center}
\includegraphics[width=3in, height=3in, angle=0]{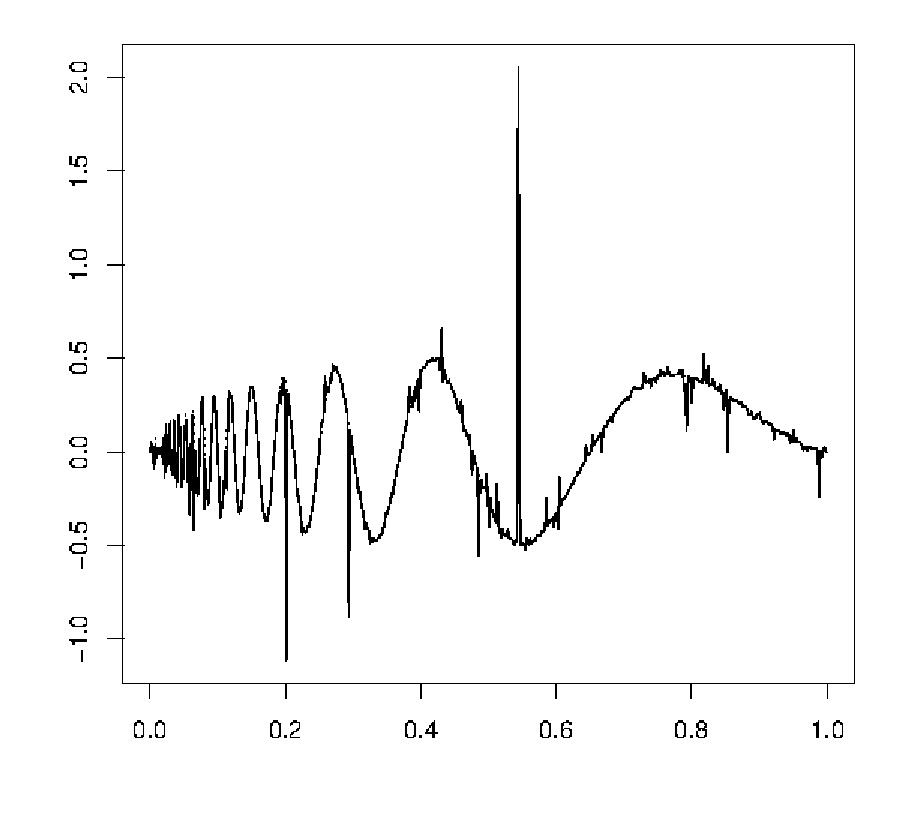}\\
\caption{The Doppler test function with Cauchy noise added to represent a ``SNR'' of 5.}
\label{figure:TestFuncsCauchy}
\end{center}
\end{figure}

When considering the type of noise to add on top of our test functions, we intentionally chose distributions with finite variance. This way the SNR is defined and can be used as a measure of the amount of noise in our data. Any distribution with infinite variance, such as Cauchy, cannot be properly scaled, and has arbitrarily large errors (see Figure \ref{figure:TestFuncsCauchy}).

The code and corresponding documentation to implement our method and generate the test functions used in this paper are provided as ancillary files for this article. 

\clearpage

\bibliographystyle{model2-names}
\bibliography{myrefs_rv}

\end{document}